\documentclass[twocolumn,aps,prb,superscriptaddress,longbibliography]{revtex4-2}
\usepackage{graphicx}
\usepackage{color}
\usepackage{amsmath}
\usepackage{amssymb}
\usepackage[normalem]{ulem}
\usepackage{adjustbox}
\usepackage{nicefrac,physics}
\usepackage{caption,subcaption,mwe}
\usepackage{ragged2e}

\renewcommand{\vec}[1]{{\boldsymbol #1}}
\newcommand{\rev}[1]{{{#1}}} 
\def\nn{\nonumber\\}

\begin{document}
\title{Kondo effect in twisted bilayer graphene}

\author{A. S. Shankar}
\affiliation{Instituut-Lorentz, Universiteit Leiden, P.O. Box 9506, 2300 RA Leiden, The Netherlands}
	
\author{D. O. Oriekhov}
\affiliation{Instituut-Lorentz, Universiteit Leiden, P.O. Box 9506, 2300 RA Leiden, The Netherlands}
	
\author{Andrew K. Mitchell}\email[]{Andrew.Mitchell@UCD.ie}
\affiliation{School of Physics, University College Dublin, Belfield, Dublin 4, Ireland}
\affiliation{Centre for Quantum Engineering, Science, and Technology, University College Dublin, Ireland}
	
\author{L. Fritz}\email[]{L.Fritz@uu.nl}
\affiliation{Institute for Theoretical Physics, Utrecht University, Princetonplein 5, 3584 CC Utrecht, Netherlands}
	
%##########################
%##########################

\begin{abstract}
\noindent The emergence of flat bands in twisted bilayer graphene at the magic angle can be understood in terms of a vanishing Fermi velocity of the Dirac cone. This is associated with van Hove singularities approaching the Fermi energy and becoming higher-order. In the density of states this is reflected by flanking logarithmic van Hove divergences pinching off the central Dirac cone in energy space. The low-energy pseudogap of the Dirac cone away from the magic angle is replaced by a power-law divergence due to the higher-order van Hove singularity at the magic angle.
 This plays an important role in the exotic phenomena observed in this material, such as superconductivity and magnetism, by amplifying electronic correlation effects.
Here we investigate one such correlation effect -- the Kondo effect due to a magnetic impurity embedded in twisted bilayer graphene. We use the Bistritzer-MacDonald model to extract the low-energy density of states of the material as a function of twist angle, and study the resulting quantum impurity physics using perturbative and numerical renormalization group methods. Although at zero temperature the impurity is only Kondo screened precisely at the magic angle, we find highly nontrivial behavior at finite temperatures relevant to experiment, due to the complex interplay between Dirac, van Hove, and Kondo physics.
\end{abstract}
\maketitle
	
%##########################
%##########################

\section{Introduction}

The properties of two-dimensional monolayer systems are strongly modified by stacking two layers with a relative twist, the so-called moir\'e effect in twisted bilayer systems~\cite{LopesdosSantos2007PRL}. In particular, twisted bilayer graphene (TBG) exhibits peculiar properties at specific `magic' twist angles~\cite{Suarez2011,Bistritzer2011,Kim2017,Liu2019,Yuan2019,Song2019,Tarnopolsky2019,Cao2018,Caocorrelated2018,Lu2019}. One characteristic of the system at these magic angles is that the non-interacting band structure contains almost-flat bands. This leads to a dramatic enhancement of the density of states (DoS). Consequently, electronic interaction effects are boosted, favoring the appearance of magnetism and other correlated phases, for example superconductivity~\cite{Cao2018,Caocorrelated2018,Lu2019,Ojajarvi2018,Isobe2018,Wu2018,Peltonen2018,Kozii2019,Yankowitz2019,Hazra2019,Hu2019,Xie2020,Julku2020}. 
Away from the magic angle in TBG, the slightly modified Dirac cones of the single layers persist~\cite{LopesdosSantos2007PRL}, giving a low-energy linear pseudogap DoS. However, pronounced van Hove singularities~\cite{vanHove1953} (vHs's) dominate the band structure at higher energies, leading to logarithmic divergences in the DoS. Experimentally, it was recently shown~\cite{Li2010} that as the magic angle is approached, the vHs divergences in the DoS move to lower energy, pinching off the Dirac cone from either side in energy space. At the magic angle, the vHs divergences in the DoS merge, and a single higher-order vHs \cite{Yuan2019} (HO-vHs) emerges, yielding a stronger, power-law divergent DoS. This is a characteristic feature of the emergent flat bands in this system. The enhanced effect of electron correlations due to the HO-vHs in the bulk TBG material has been studied theoretically~\cite{Caocorrelated2018,Lu2019,Isobe2018,Classen2020PRB} and confirmed in scanning tunnelling spectroscopy (STS) experiments~\cite{Kerelsky2019,Choi2019}.

Detailed information on the electronic structure of new materials can also be obtained by exploiting defects or impurities as \textit{in-situ} probes \cite{crommie1993imaging}. The nature of the electronic scattering from impurities in a system is strongly dependent on the band structure and DoS of the clean host material, and can be probed either locally at the impurity site by STS \cite{jamneala2000scanning,madhavan2001local}, or by collecting momentum-space information through quasiparticle interference (QPI) measurements \cite{hoffman2002imaging}. For quantum impurities such as magnetic adatoms \cite{costi2009kondo,ternes2008spectroscopic} or single-molecule magnets \cite{bogani2008molecular}, the impurity spin degree of freedom generates additional spin-flip scattering, which is boosted at low temperatures by the Kondo effect in standard metallic hosts \cite{Hewson}. The Kondo effect itself depends sensitively on the local spin-resolved DoS of the host material, and hence such `Kondo probes' can provide additional electronic structure information \cite{derry2015quasiparticle,*mitchell2015multiple} or be utilized for quantum metrology \cite{mihailescu2022thermometry}. Aside from the spectroscopic and QPI signatures of Kondo physics in metals \cite{ternes2008spectroscopic}, the Kondo effect has been studied in a range of other unconventional materials, including monolayer graphene \cite{chen2011tunable,vojta2010gate,Fritz2013,mitchell2013kondo}, topological insulators \cite{mitchell2013TI}, Dirac and Weyl semimetals \cite{mitchell2015kondo}, ferromagnets \cite{martinek2003kondo,calvo2009kondo}, superconductors \cite{muller1971kondo,polkovnikov2001impurity}, and spin liquids \cite{kolezhuk2006theory,vojta2016kondo,he2022magnetic} -- each giving its own distinctive response. In particular, for Dirac systems with a low-energy pseudogap, the depleted conduction electron DoS is known to suppress the Kondo effect \cite{Fritz2013,mitchell2013TI,mitchell2015kondo} (although it can be revived upon doping \cite{May2018PRB-Kondo-graphene}).

By contrast, in the case of TBG, one might expect Kondo correlations to be strongly enhanced by the flat bands and diverging DoS close to the magic angle. The study of Kondo physics in TBG, and how it evolves with twist angle, is the topic of this article. We find that magnetic impurities are sensitive probes of the nontrivial band structure of the material, and we uncover rich thermodynamic and spectroscopic signatures that rapidly change on approaching the magic angle.

\begin{figure*}[t!]
    \begin{subfigure}[b]{0.275\linewidth}
        \centering
        \includegraphics[width=1.1\linewidth]{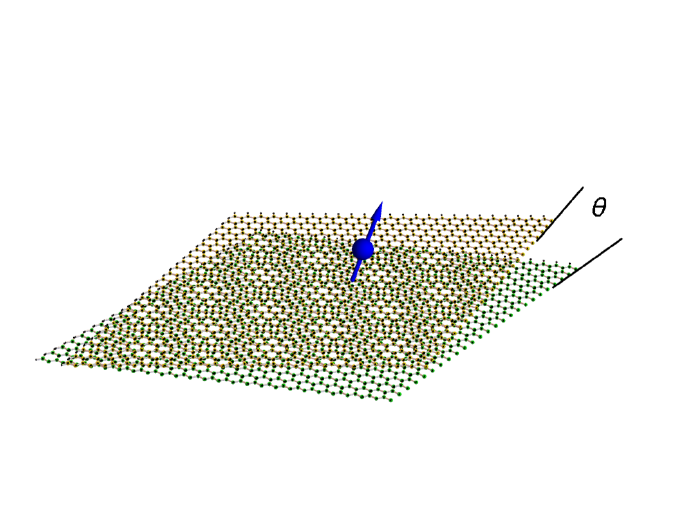}
        \caption{\label{fig:schematic}}
    \end{subfigure}
    \begin{subfigure}[b]{0.400\linewidth}
        \centering
        \includegraphics[width=\linewidth]{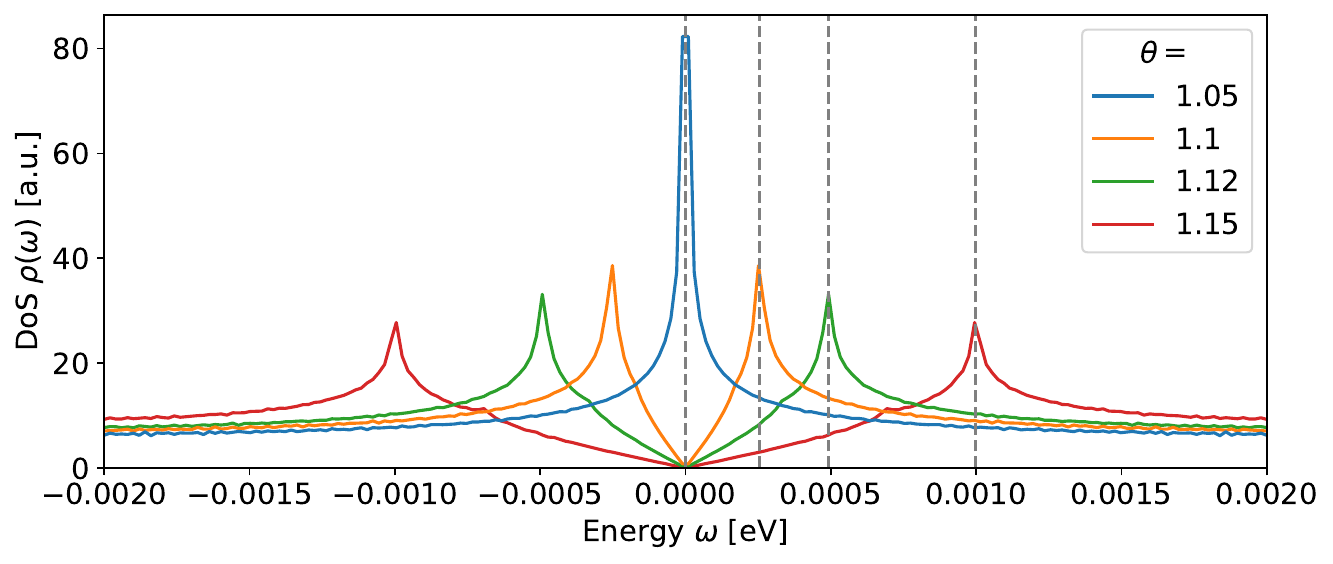}
        \caption{{\label{fig:HistogramDOS}}}
    \end{subfigure}
    \begin{subfigure}[b]{0.300\linewidth}
        \centering
        \includegraphics[width=\linewidth]{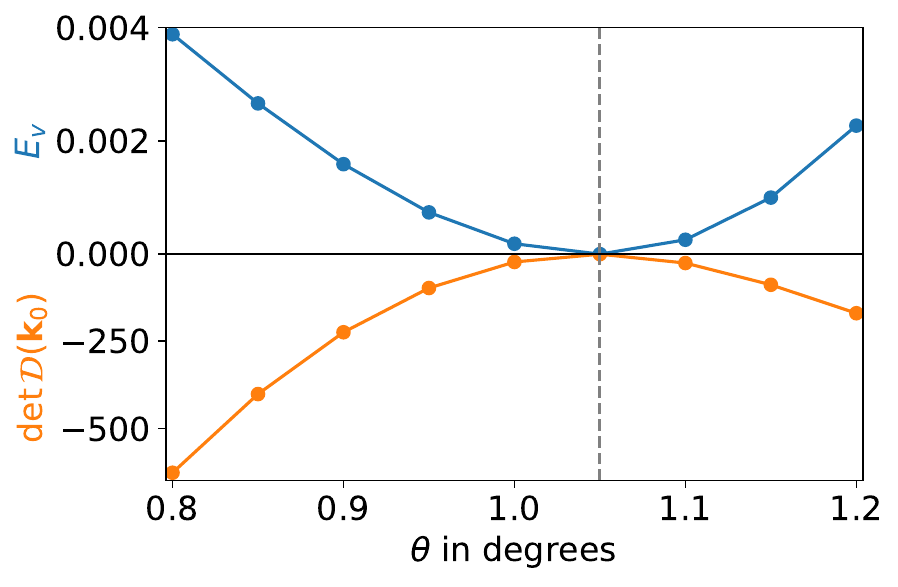}
        \caption{{\label{fig:ShowHigherOrder}}}
    \end{subfigure}
\caption{ %\justifying
\small (a) Schematic representation of a magnetic impurity on the surface of the twisted bilayer graphene host material, with inter-layer twist angle $\theta$.   (b) Evolution of the clean TBG density of states for different twist angles near the magic angle at $\theta = 1.05^{\circ}$. Vertical dashed lines indicate the energy of the dispersion saddle points $E_v$, which determine the van Hove singularity locations. (c) Vanishing vHs saddle point scale $E_v$ (top) and determinant of the saddle point Hessian (bottom) as the magic angle is approached, showing how two vHs's coalesce into a single HO-vHs. }
    \label{fig:joinedfig}
\end{figure*}

Specifically, we consider a single, interacting Anderson impurity embedded in the TBG host -- see Fig.~\ref{fig:schematic}. The clean TBG material is modelled using the Bistritzer-MacDonald (BM) model~\cite{Bistritzer2011}, which we discuss in Sec.~\ref{sec:BM-model}. We focus on the role of the vHs's and their evolution with twist angle. The model shows an intricate interplay between different DoS elements: metallic, Dirac pseudogap, vHs log-divergence, and HO-vHs power-law divergence. In Sec.~\ref{sec:Kondo} we review the physics of the Kondo model, emphasizing the different limiting behaviors arising in the metallic, pseudogap, and log-diverging or power-law diverging DoS needed to understand the compound DoS structure in TBG. Finally, in Sec.~\ref{sec:Kondo-BM} we present full numerical renormalization group (NRG) results for an Anderson impurity in TBG in the vicinity of, and at, the magic angle. We focus on thermodynamic quantities such as the impurity entropy as a clear means of identifying the different fixed points and emergent energy scales. We furthermore study the energy-dependence of the local impurity spectral function, which is relevant to STS experiments. We conclude in Sec.~\ref{sec:conclusions}, commenting on the suitability of magnetic impurities as \textit{in-situ} probes for the physics of TBG near the magic angle, and an outlook for experiments. Technical material is given for reference in the Appendices. 

We note that the Kondo model we consider in this work is completely different from recent studies of TBG as a heavy fermion problem~\cite{song2022magic,hu2023kondo,hu2023symmetric,zhou2023kondo}, where the quenched kinetic energy of the flat band lends itself to being treated as an immobile lattice of impurities. In those works, the correlated local moments are a part of the TBG lattice itself, whereas here we consider additional adatom impurities coupled to the TBG host. The effective impurity models and corresponding electronic hybridization functions are rather different in these two cases.

%##########################
%##########################

\section{Van Hove singularities in the Bistritzer-MacDonald model} \label{sec:BM-model}

Before considering a Kondo impurity in TBG, we first analyze the clean host material, focusing on how the vHs's affect the band structure and local DoS. In the first part of this section we briefly recall the details of the Bistritzer-MacDonald (BM) model of TBG and its particle-hole symmetric limit. The original derivations were performed in Refs.~\cite{Bistritzer2011,Suarez2011,Bernevig2019PRL-TBG}; further details are provided in the Appendices. In the second part, we analyze the formation of flat bands from saddle points and discuss the detailed structure of the lowest energy bands.
	
%################

\subsection{Particle-hole symmetric\\Bistritzer-MacDonald model}

To describe TBG with a small twist angle  $\theta$, it is necessary to take into account both the intralayer hopping parameter $t$ for each of the individual graphene layers, as well as the interlayer tunneling $w$. In the following we take these to be $t\approx 2.87 \,\text{eV}$ and $w\approx 0.11 \,\text{eV}$, as used in Ref.~\cite{Bistritzer2011}. The twist angle between the layers generates a Moir\'{e} pattern with an emergent superlattice structure. For small twist angles $\theta$, the characteristic Moir\'{e} length scale $L_{\theta}$ is given by $L_{\theta}=\sqrt{3} a /[2 \sin (\theta / 2)]$ with $a=1.42\text{\AA}$ being the interatomic distance in monolayer graphene. The corresponding effective low-energy Hamiltonian near the $K$ point of the Moir\'{e} Brillouin zone (MBZ) has the form \cite{Bistritzer2011,Bernevig2019PRL-TBG},
\begin{align}
\label{eq:H-TBG-full}
&H(\vec{k})=\nn
&\hspace{-0.3cm}\left(\begin{array}{cccc}
h_{\frac{\theta}{2}}^{K}(\vec{k}) & w T_1 & w T_2 & w T_3 \\
w T_1^{\dagger} & h_{-\frac{\theta}{2}}^{K}(\vec{k}-\mathbf{q}_1) & 0 & 0 \\
w T_2^{\dagger} & 0 & h_{-\frac{\theta}{2}}^{K}(\vec{k}-\mathbf{q}_2) & 0 \\
w T_3^{\dagger} & 0 & 0 & h_{-\frac{\theta}{2}}^{K}(\vec{k}-\mathbf{q}_3)
\end{array}\right).
\end{align}
The wave vector $\vec{k}$ is measured relative to the $K$ point, and the Hamiltonian acts on 8-component wavefunctions $\Psi=\left(\psi_{0, \vec{k}}, \psi_{1, \vec{k}}, \psi_{2, \vec{k}}, \psi_{3, \vec{k}}\right)^T$, where $\psi_{0, \vec{k}}$ is a two-component spinor in the A-B sublattice basis in the top layer, and $\psi_{1(2,3), \vec{k}}$ are spinors in bottom layer at wave vectors $\vec{k}-\vec{q}_{1(2,3)}$.
Here $h_{\phi}^{K}(\vec{k})$ is the effective low-energy Hamiltonian of single layer graphene near the $K$ point, in a coordinate frame rotated by angle $\phi$,
\begin{align}
h_{\phi}^K(\vec{k})=k v_F \left[\begin{array}{cc}
0 & e^{i\left(\theta_{\vec{k}}-\phi\right)} \\
e^{-i\left(\theta_{\vec{k}}-\phi\right)} & 0
\end{array}\right]\;.
\label{eq:hKtheta}
\end{align}
Here, the angle $\theta_k$ measures the orientation of the momentum relative to
the $x$-axis, $k=|\vec{k}|$, and the Fermi velocity is $v_F=9.3 \times 10^7 \mathrm{~cm} / \mathrm{s}$. The wave vectors $\vec{q}_{1,2,3}$ connecting $K$-points of the top and bottom layers are,
\begin{subequations}
\begin{align}
&\mathbf{q}_1=k_\theta\{0,-1\}, \\
&\mathbf{q}_2=k_\theta\left\{\frac{\sqrt{3}}{2}, \frac{1}{2}\right\}, \\
&\mathbf{q}_3=k_\theta\left\{-\frac{\sqrt{3}}{2}, \frac{1}{2}\right\},
\end{align} 
\end{subequations}
with Moir\'{e} wave number, $k_\theta \equiv |\mathbf{q}_j|=\frac{8 \pi}{3 \sqrt{3} a} \sin \left(\frac{\theta}{2}\right)$.

Finally, the interlayer tunneling matrices $T_{1(2,3)}$ are expressed in terms of Pauli matrices, viz:
\begin{subequations}
\begin{align}
&T_1=1+\sigma_x \;,\\
&T_2=1-\frac{\sigma_x}{2}-\frac{\sqrt{3}\sigma_y}{2} \;,\\
&T_3=1-\frac{\sigma_x}{2} +\frac{\sqrt{3}\sigma_y}{2}  \;.
\end{align}
\end{subequations}

The Hamiltonian Eq.~\eqref{eq:H-TBG-full} captures the essential physics of TBG and correctly predicts the first magic angle at $\theta\simeq 1.05^{\circ}$. At the $K$ point of the MBZ one finds that the lowest energy bands have a Dirac cone dispersion with an effective Fermi velocity,
\begin{align}
v_F^{\star}=\frac{1-3 w^2 /\left(\hbar v_F k_\theta\right)^2}{1+6 w^2 /\left(\hbar v_F k_\theta\right)^2}
\end{align}
which vanishes exactly at the magic angle, where $\hbar v_F k_\theta=\sqrt{3} w$.

In the following, we will use the particle-hole symmetric version of the BM model. This form is obtained from Eq.~\eqref{eq:H-TBG-full} by eliminating subleading (second-order) corrections in the diagonal elements coming from the effect of the twist on the single layer Hamiltonian \cite{Bernevig2019PRL-TBG}. This is simply achieved by setting $\phi=0$ in Eq.~\eqref{eq:hKtheta},
\begin{align}
	\label{eq:H-TBG-symmetric}
H(\vec{k})=\left(\begin{array}{cccc}
h_0^{K}(\vec{k}) & w T_1 & w T_2 & w T_3 \\
w T_1^{\dagger} & h_0^{K}(\vec{k}-\mathbf{q}_1) & 0 & 0 \\
w T_2^{\dagger} & 0 & h_0^{K}(\vec{k}-\mathbf{q}_2) & 0 \\
w T_3^{\dagger} & 0 & 0 & h_0^{K}(\vec{k}-\mathbf{q}_3)
\end{array}\right)\;.
\end{align}

The low-energy TBG DoS $\rho(\omega)$, obtained by diagonalizing this Hamiltonian is shown in Fig.~\ref{fig:HistogramDOS}, for different twist angles in the vicinity of the magic angle at $\theta=1.05^{\circ}$. When precisely at the magic angle, we see a single divergence in the DoS at the Fermi energy. We measure energies relative to the Fermi energy and set $E_F=0$, such that $\rho(\omega)=\rho(-\omega)$, embodying the particle-hole symmetry of Eq.~\eqref{eq:H-TBG-symmetric}. However, moving away from the magic angle, we have a Dirac cone feature, with pseudogap vanishing DoS $\rho(\omega)\sim |\omega|$ below an emergent scale $|\omega|\ll E_v$. We also see two vHs points with diverging DoS at $\omega=\pm E_v$. Below we analyze the low-energy bands and vHs structure of the model, extracting the angle dependence of the vHs scale $E_v$.

%######################
	
\subsection{Characterization of van Hove singularities in the particle-hole symmetric BM model}

We start our analysis of vHs properties in TBG by recalling the classification recently introduced in Refs.~\cite{Yuan2019,Chamon2020PRR,Yuan2020PRB-classification}, which expands the definition from the usual vHs with logarithmically diverging DoS  \cite{vanHove1953} to include HO-vHs with power-law diverging DoS. For a band dispersion $\epsilon(\vec{k})$, which is a function of the 2D momentum vector $\vec{k}$, we calculate the first derivatives $\nabla_{\vec{k}}\epsilon(\vec{k})$, and the Hessian matrix of second derivatives $\mathcal{D}_{i j}(\vec{k}) \equiv \frac{1}{2} \partial_{k_i} \partial_{k_j} \varepsilon(\vec{k})$. The Hellman-Feynman theorem allows to carry this out with high numerical accuracy (see Appendix A). 
A logarithmic vHs arises at a point $\vec{k}_0$ in the dispersion corresponding to a saddle point when:
\begin{align}\label{eq:logvhs}
\text{log vHs:}\quad \nabla_{\vec{k}}  \varepsilon(\vec{k}_0)=\mathbf{0} \text {~~~and~~} \operatorname{det} \mathcal{D}(\vec{k}_0)<0 \;.
\end{align}
The negative Hessian determinant means that we have both a maximum and a minimum in each of the principal directions of the saddle point. A higher-order saddle point, corresponding to a HO-vHs, is instead characterized by zero determinant of Hessian matrix:
\begin{align}\label{eq:hovhs}
\text{HO-vHs:}\quad \nabla_{\vec{k}}  \varepsilon(\vec{k}_0)=\mathbf{0} \text {~~~and~~} \operatorname{det} \mathcal{D}(\vec{k}_0)=0 \;.
\end{align}
In addition, higher-order saddle points can be classified according to the leading polynomial terms in an expansion of the dispersion $\epsilon(\vec{k})$ around the saddle point $\vec{k}_0$ \cite{Chamon2020PRR,Yuan2020PRB-classification}. These leading terms in the expansion correspond directly to the numerical exponent of the power-law divergence in the DoS.

	\begin{figure*}
        \centering
        \begin{subfigure}[b]{0.475\linewidth}
            \centering
            \includegraphics[width=\linewidth]{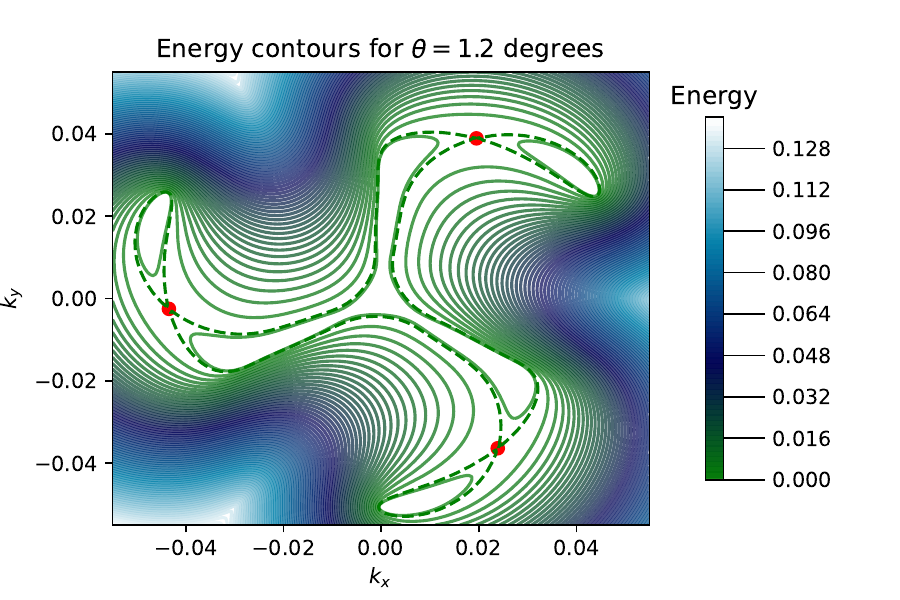}
            \caption[ContoursLog]%
            {{\small Conventional  vHs}\label{fig:contourslog}}  
        \end{subfigure}
        \hfill
        \begin{subfigure}[b]{0.475\linewidth}   
            \centering 
            \includegraphics[width=\linewidth]{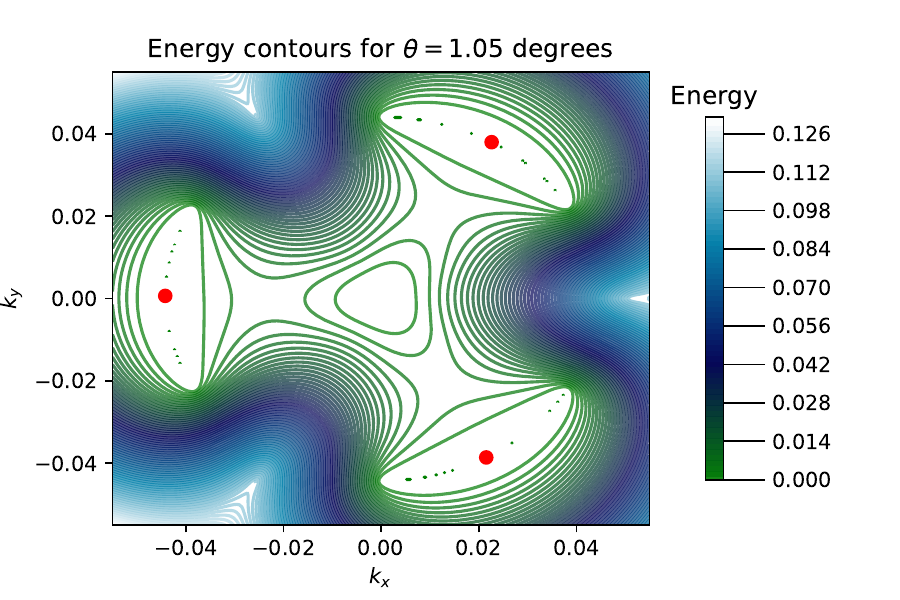}
            \caption[]%
            {{\small Higher Order vHs at the magic angle}\label{fig:Magic_Heatmap}}    
        \end{subfigure}
        \caption[ FermiSurfacePlots ]
        {%\justifying
        \small  Low-energy features of the particle-hole symmetric BM model of pristine TBG, showing energy contours of the dispersion $\varepsilon(\boldsymbol{k})$ as a function of momentum $\boldsymbol{k}$ in the extended MBZ. Red points show the saddle points of the dispersion, corresponding to the van Hove singularities. Dashed lines show the contours at energy $E_v$, on which the van Hove points sit.} 
        \label{fig:Spinners}
    \end{figure*}

With the help of this classification, we proceed to investigate the structure of the lowest band in the BM model Hamiltonian given by Eq.~\eqref{eq:H-TBG-symmetric}. Using particle-hole symmetry in conjunction with the transformation $k_x\to -k_x$ allows us to concentrate only on the lowest positive energy band for our analysis. Technical details of our numerical implementation are presented in  Appendix A. In Fig.~\ref{fig:ShowHigherOrder} we show the numerically-computed $E_v$ scale (top panel) and the determinant of the saddle-point Hessian $\operatorname{det} \mathcal{D}(\vec{k}_0)$ (bottom panel) as a function of twist angle. Both are seen to vanish at the magic angle, heralding the emergence of the HO-vHs at this point. This behavior is further analyzed below.

In Fig.~\ref{fig:Spinners} we plot the momentum-resolved spectrum of the BM model for twist angles $\theta=1.2^{\circ}$ (panel a) and $\theta=1.05^{\circ}$ (panel b). We compute $\nabla_{\vec{k}} \varepsilon(\vec{k})$ throughout the MBZ and search for the vHs points for which $\nabla_{\vec{k}} \varepsilon(\vec{k})=\vec{0}$. These are indicated in both panels as the red circle points. 

Having located the vHs points in the MBZ for a given twist angle, we can classify them and study their neighborhood in momentum space. Away from the magic angle (e.g.~for $\theta=1.2^{\circ}$ shown in Fig.~\ref{fig:Spinners}a), we indeed find a negative determinant of the Hessian $\operatorname{det} \mathcal{D}(\vec{k}_0)<0$, and the vHs's have a local hyperbolic geometry in momentum space. The dispersion is found to have the leading form,
\begin{equation}
\epsilon_{\mathbf{p}} = E_v + \alpha p_x^2 - \beta p_y^2 \;, 
\label{eq:Logdispersion}
\end{equation}
where the coefficients are obtained from the eigenvalues of the Hessian, and the labels $p_x$ and $p_y$ are measured in the principal directions of the saddle point, with $\mathbf{p}=0$ defining the saddle point itself. The corresponding leading correction to the DoS then takes the form,
	\begin{equation}
	    \rho(\omega) = \frac{1}{4\pi^2}\frac{1}{\sqrt{\alpha\beta}}\ln \abs{\frac{D}{\omega-E_v}} \;,
	    \label{eq:logDOS}
	\end{equation}
where $D$ is a high-energy cutoff, taken to be the conduction electron bandwidth. In this way, we may extract the vHs scale $E_v$.

As we begin twisting toward the magic angle, the vHs energy scale $E_v$ starts to decrease, and the two logarithmic singularities at $\omega=\pm E_v$ therefore move closer together. Furthermore, the magnitude of the Hessian at the vHs, $|\operatorname{det} \mathcal{D}(\vec{k}_0)|$, also decreases. The two vHs points merge at $\omega=0$ precisely at the magic angle $\theta = 1.05^\circ$ \cite{Li2010}, at which point the Hessian also vanishes, $\operatorname{det} \mathcal{D}(\vec{k}_0)=0$. This transition is shown in Fig.~\ref{fig:ShowHigherOrder}. As the magic angle is approached and the HO-vHs is formed, we see a further flattening of the dispersion in the $p_y$ direction. The fittingly-named higher-order singularity requires a higher-order polynomial to faithfully capture its dispersion. The lowest polynomial which correctly captures all the symmetries is given by,
	\begin{equation}
	   \epsilon_{\mathbf{p}} = \alpha p_x^2 + \gamma p_x p_y^2 + \kappa p_y^4 \;.
	\end{equation}
This comes hand-in-hand with a sharper, power-law divergence in the DoS, 
\begin{equation}\label{eq:HOdos}
\rho(\omega) = (2\pi)^{-\frac{5}{2}} \Gamma(\tfrac{1}{4})^2 (4\alpha\Tilde{\Gamma}^2)^{-1/4}~ |\omega|^{-\nicefrac{1}{4}} \;, 
\end{equation}
with $\Tilde{\Gamma}^2=\gamma^2 - 4\alpha\kappa $ and $\Gamma(x)$ the usual gamma function. 
	
The low-energy DoS at different angles can be computed numerically by binning histograms of energies for the lowest band of the BM model, as shown in Fig.\ref{fig:HistogramDOS}. Away from the magic angle, the numerical calculation indicates a linear DoS around the Fermi energy, coming from the Dirac cone in the spectrum. Around $\omega=\pm E_v$ we see the vHs log-divergences. At the magic angle, the HO-vHs around the Fermi energy `eat up' the Dirac cone, and we have instead a large DoS at low energies, diverging as $\rho(\omega)\sim |\omega|^{-1/4}$.

%#####################
%#####################

\section{The Kondo problem}\label{sec:Kondo}

The Kondo effect is a classic paradigm in many-body quantum physics \cite{Hewson}. The corresponding Kondo model features a single quantum spin-$\tfrac{1}{2}$ magnetic impurity coupled by antiferromagnetic exchange to a single channel of non-interacting conduction electrons. Originally, the Kondo model was formulated to describe dilute magnetic impurities such as iron, in bulk metals such as gold \cite{Kondo1970,costi2009kondo}. In these metallic systems, an impurity local moment becomes strongly entangled with its surrounding conduction electrons at low temperatures, and is dynamically screened \cite{bayat2010negativity,mitchell2011real,v2020observation}. This leads to dramatically enhanced spin-flip scattering in the host metal, which can be detected spectroscopically \cite{ternes2008spectroscopic}. 

A more microscopic starting point is provided by the single impurity Anderson model \cite{Hewson}, which describes the impurity as a single quantum orbital with strong electron interactions,
\begin{eqnarray}\label{eq:AM}
    H_{\rm AM} = H_{\rm host} &+ \epsilon_d \sum_{\sigma} d_{\sigma}^{\dagger}d_{\sigma}^{\phantom{\dagger}} + U_d d_{\uparrow}^{\dagger}d_{\uparrow}^{\phantom{\dagger}}d_{\downarrow}^{\dagger}d_{\downarrow}^{\phantom{\dagger}} \nonumber \\ &+ g\sum_{\sigma}\left ( d_{\sigma}^{\dagger} c_{0,\sigma}^{\phantom{\dagger}} + c_{0,\sigma}^{\dagger}d_{\sigma}^{\phantom{\dagger}} \right) \;,
\end{eqnarray}
where $d_{\sigma}^{(\dagger)}$ is an annihilation (creation) operator for an impurity electron with spin $\sigma=\uparrow$/$\downarrow$, and $H_{\rm host}=\sum_{k,\sigma} \epsilon_k c_{k,\sigma}^{\dagger}c_{k,\sigma}^{\phantom{\dagger}}$ describes the clean host. Here $c_{k,\sigma}^{(\dagger)}$ annihilates (creates) a conduction electron of the host material with momentum $k$ and spin $\sigma$. We do not employ band indices in this expression. The impurity couples locally in real-space to the effective host orbital $c_{0,\sigma}=\sum_k \xi_k c_{k,\sigma}$, where $\xi_k$ is the weight of state $k$ at the impurity location (taken to be at the origin).

For small host-impurity hybridization $g$, repulsive Coulomb interaction $U_d>0$, and suitably-chosen impurity potential $-U_d<\epsilon_d<0$, a spin-$\tfrac{1}{2}$ local moment can be trapped on the impurity. Projecting onto this doubly-degenerate spin-$\tfrac{1}{2}$ manifold of impurity states by eliminating virtual excitations to empty or doubly-occupied impurity configurations by means of a Schrieffer-Wolff (SW) transformation \cite{Hewson,schrieffer1966relation} yields the simpler Kondo model,
\begin{eqnarray}\label{eq:KM}
    H_{\rm KM} = H_{\rm host} &+ J \vec{S} \cdot \vec{s}_0 + V\sum_{\sigma}c_{0,\sigma}^{\dagger}c_{0,\sigma}^{\phantom{\dagger}}  \;,
\end{eqnarray}
where $J>0$ is the antiferromagnetic exchange interaction between the impurity spin-$\tfrac{1}{2}$ degree of freedom $\vec{S}$, and the \emph{spin density} of the host conduction electrons at the impurity position $\vec{s}_0=\tfrac{1}{2}\sum_{\alpha,\beta}c^{\dagger}_{0,\alpha} \vec{\sigma}_{\alpha\beta} c^{\phantom{\dagger}}_{0,\beta}$, where $\vec{\sigma}$ is the vector of Pauli matrices. The third term describes potential scattering of the host conduction electrons induced by the impurity (since $c_{0,\sigma}^{\dagger}c_{0,\sigma}^{\phantom{\dagger}}=\sum_{k,k'}\xi_k\xi_{k'}c_{k,\sigma}^{\dagger}c_{k',\sigma}^{\phantom{\dagger}}$). 

The standard Schrieffer-Wolff result \cite{Hewson,schrieffer1966relation}, which becomes exact in the limit $U_d/g^2\to \infty$, yields $J=2g^2[(U_d+\epsilon_d)^{-1} - (\epsilon_d)^{-1}]$ and $V=-g^2[(U_d+\epsilon_d)^{-1} + (\epsilon_d)^{-1}]$. At the particle-hole symmetric point of the model $\epsilon_d=-U_d/2$, we therefore obtain $J=8g^2/U_d$ and $V=0$ (the latter result can be viewed as a many-body quantum destructive interference effect between particle and hole processes). Although the Kondo model Eq.~\eqref{eq:KM} correctly captures the low-energy physics of Eq.~\eqref{eq:AM}, it should be noted that for realistic values of $U_d$, $\epsilon_d$, and $g$ outside of the strict perturbative regime, the values of the effective parameters $J$ and $V$ must be obtained by more sophisticated means that take into account renormalization from the conduction electrons and the specific conduction electron DoS \cite{rigo2020machine}. We also emphasize that $J$ and $V$ are not independent parameters in Eq.~\eqref{eq:KM}, being both derived from the same microscopic parameters of the underlying Anderson model.

The physics of the impurity problem is controlled by the local (free) conduction electron DoS seen by the impurity, $\rho_{\sigma}(\omega)=-\tfrac{1}{\pi}{\rm Im}~\langle\langle c_{0,\sigma}^{\phantom{\dagger}} ; c_{0,\sigma}^{\dagger}\rangle\rangle^{0}$, where $\langle\langle c_{0,\sigma}^{\phantom{\dagger}} ; c_{0,\sigma}^{\dagger}\rangle\rangle$ is the retarded, real-frequency local host Green's function at the impurity position, and the $0$ superscript denotes that it is calculated for the clean host.

In this work we consider a single magnetic impurity (the `dilute limit') embedded in an otherwise clean host TBG system modelled by the BM model. $SU(2)$ spin symmetry is taken to be unbroken. We also assume that the impurity couples equally to all BM bands independently of the impurity position ($\xi_k$ is constant for all momenta and band indices), such that $\rho_{\sigma}(\omega)\equiv \rho(\omega)$ is the TBG DoS, whose low-energy form is described by Eqs.~\eqref{eq:logDOS} or \eqref{eq:HOdos}. \rev{This is certainly a simplification, since details of the impurity-TBG hybridization will naturally affect details of the impurity response. The specific form of the impurity hybridization function will in practice depend on the impurity location within the moir\'e unit cell and how the impurity couples in real-space to the constituent TBG carbon atoms. We leave such \textit{ab initio} studies for future work. However, the rich physics uncovered below will remain qualitatively unaltered provided the impurity hybridization function still features  van Hove divergences flanking a central pseudogap Dirac cone. Since the origin of these features is rooted deeply in the symmetry and topology of the TBG material, we expect the idealized Kondo physics described here to be rather generic. On the other hand, insights from monolayer graphene \cite{mitchell2013kondo} indicate that the physics of vacancies in TBG or substitutional dopants may drastically differ, since the local DoS in these cases is strongly modified.}

In the rest of this section we review the methods that we use to attack the problem as well as the quantities to be analyzed. 

%################+

\subsection{Poor man's scaling approach}\label{subsec:poorman}

The Kondo model as defined by Eq.~\eqref{eq:KM} is a nontrivial strong correlation problem. For metallic host systems, the first insights were provided by Kondo's calculation of the scattering T-matrix \cite{Kondo1970}, which is related to the impurity spectral function. Kondo found a low-temperature divergence in perturbation theory: even when the bare $J$ is small, straight perturbation theory does not give a good description of the low-temperature physics or a proper understanding of the many-body ground state of the system. This divergence was better understood by Anderson's self-coined `poor-man's scaling' approach \cite{anderson1970poor} -- a precursor to the renormalization group (RG). It identified an emergent low-energy scale $T_K$ -- the Kondo temperature -- below which perturbation theory breaks down, and the problem becomes a strong coupling problem. We briefly introduce the method here, since we will employ it in the next section to understand analytically the scaling properties of an impurity in the TBG host. 

Conventionally in the poor man's scaling approach, one uses dimensionless couplings $j=\rho_0 J$ and $v=\rho_0 V$ with $\rho_0=\rho(\omega=0)$ the Fermi level DoS. However, for consideration of Dirac systems where $\rho_0$ may in fact vanish, a different choice is required. Here we simply use the original dimensionful couplings $J$ and $V$. Furthermore, we assume that the host DoS is particle-hole symmetric, meaning $\rho(\omega)=\rho(-\omega)$, a property satisfied by Eq.~\eqref{eq:H-TBG-symmetric}; and is defined within a band of halfwidth $D$, meaning $\rho(\omega)\propto \theta(D-|\omega|)$.

Anderson's scaling procedure goes as follows \cite{anderson1970poor}: (i) integrate out high energy conduction electron states $D-\delta D<|\omega|<D$ close to the band edges in a shell of width $\delta D$; (ii) incorporate the effect of virtual excitations to these eliminated states perturbatively by rescaling the couplings $J$ and $V$ to give an effective Hamiltonian of the same form but defined with a reduced bandwidth $D\to D-\delta D$; (iii) study the flow of the parameters $J$ and $V$ on successive reduction of the bandwidth. Making $\delta D$ infinitessimal, one obtains the following scaling equations,
\begin{eqnarray}\label{eq:poorman}
\frac{dJ}{d D}&=&-\frac{\rho(D)}{D}J^2 \;, \nonumber \\ 
\frac{dV}{d D}&=&0\;.
\end{eqnarray}
The first equation determines the flow of the coupling constant $J$ on reducing the bandwidth, whereas the second equation shows that the potential scattering $V$ does not flow. If the bare model is particle-hole symmetric then no potential scattering is generated under the scaling procedure. In the remainder of this paper we will focus on the case $V=0$. The scaling equation for $J$ gives insight into the breakdown of perturbation theory and hence $T_K$, by identifying the point where the rescaled $J$ diverges. We consider various relevant situations in the following. 

%################

\subsection{Numerical Renormalization Group}\label{subsec:nrg}

Wilson's numerical renormalization group \cite{wilson1975renormalization,bulla2008numerical} (NRG) is a non-perturbative technique for solving quantum impurity type problems. It builds upon Anderson's perturbative scaling ideas \cite{anderson1970poor}, but overcomes its limitations by establishing a more general framework in terms of which physical quantities can be calculated numerically-exactly, down to zero temperature. Wilson's original formulation of NRG \cite{wilson1975renormalization}, designed to obtain the thermodynamic properties of a single magnetic impurity in a metal, has since been extended to deal with arbitrary host systems \cite{chen1995kondo,*bulla1997anderson,bulla2008numerical}, and to the calculation of dynamical quantities via the full-density-matrix NRG approach \cite{anders2006spin,weichselbaum2007sum}. The former has allowed NRG to be applied to monolayer graphene \cite{vojta2010gate}, and other Dirac systems \cite{mitchell2013TI, mitchell2015kondo}. The latter provides access to highly accurate spectral data, with excellent real-frequency resolution, at any temperature. NRG has also been adapted over the years to extend the range of problems that can be tackled and improve accuracy or efficiency \cite{bulla1998numerical,oliveira1994generalized,bulla2003numerical,pruschke2009energy,mitchell2014generalized,*stadler2016interleaved,lee2021computing,rigo2022automatic}, making it the gold-standard method of choice for solving generalized quantum impurity problems.

The basic NRG algorithm \cite{wilson1975renormalization,bulla2008numerical} proceeds as follows. 
(i) The local conduction electron density of states $\rho(\omega)$ of the pristine host material (without the impurity) must first be calculated. \\
(ii) This DoS is then discretized logarithmically by dividing it up into intervals according to the discretization points $\pm D \Lambda^{-n}$, where $D$ is the bare conduction electron bandwidth, $\Lambda>1$ is the NRG discretization parameter, and $n=0,1,2,3,...$. The continuous electronic density in each interval is replaced by a single pole at the average position with the same total weight, yielding $\rho^{\rm disc}(\omega)$.\\
(iii) The conduction electron part of the Hamiltonian $H_{\rm host}$ is then mapped into the form of a `Wilson chain',
\begin{eqnarray}
    H_{\rm host} \to H_{\rm host}^{\rm disc} = \sum_{\sigma}\sum_{n=0}^{\infty} \Big [ &t_n^{\phantom{\dagger}} \left ( f_{n,\sigma}^{\dagger}f_{n+1,\sigma}^{\phantom{\dagger}}+  f_{n+1,\sigma}^{\dagger}f_{n,\sigma}^{\phantom{\dagger}}\right ) \nonumber \\ 
    &+\epsilon_n^{\phantom{\dagger}} f_{n,\sigma}^{\dagger}f_{n,\sigma}^{\phantom{\dagger}} \Big ]\;,\qquad
\end{eqnarray}
where the Wilson chain coefficients $\{t_n\}$ and $\{\epsilon_n\}$ are determined such that the DoS at the end of the chain reproduces exactly the discretized host DoS, that is $-\tfrac{1}{\pi}{\rm Im}~\langle\langle f_{0,\sigma}^{\phantom{\dagger}} ; f_{0,\sigma}^{\dagger}\rangle\rangle = \rho^{\rm disc}(\omega)$. For a system with particle-hole symmetry, $\epsilon_n=0$ for all $n$. Due to the logarithmic discretization, the Wilson chain hopping parameters decay roughly exponentially down the chain, $t_n \sim \Lambda^{-n/2}$, although the precise details are also important since they encode the specific host DoS \cite{bulla2008numerical}.\\ 
(iv) The impurity is coupled to site $n=0$ of the Wilson chain. We define a sequence of Hamiltonians $H_N$ comprising the impurity and the first $N$ Wilson chain sites,
\begin{equation}\label{eq:Hn}
\begin{split}
    H_N=&H_{\rm imp} + H_{\rm hyb} + \sum_{\sigma}\Bigg [ \sum_{n=0}^{N} 
    \epsilon_n^{\phantom{\dagger}} f_{n,\sigma}^{\dagger}f_{n,\sigma}^{\phantom{\dagger}}\\
        &+ \sum_{n=0}^{N-1}  t_n^{\phantom{\dagger}} \left ( f_{n,\sigma}^{\dagger}f_{n+1,\sigma}^{\phantom{\dagger}}+  f_{n+1,\sigma}^{\dagger}f_{n,\sigma}^{\phantom{\dagger}}\right ) \Bigg ] \;,
\end{split}
\end{equation}
where for the Anderson model $H_{\rm imp} =\epsilon_d \sum_{\sigma} d_{\sigma}^{\dagger}d_{\sigma}^{\phantom{\dagger}} + U_d d_{\uparrow}^{\dagger}d_{\uparrow}^{\phantom{\dagger}}d_{\downarrow}^{\dagger}d_{\downarrow}^{\phantom{\dagger}}$ and $ H_{\rm hyb}= g\sum_{\sigma} ( d_{\sigma}^{\dagger} f_{0,\sigma}^{\phantom{\dagger}} + f_{0,\sigma}^{\dagger}d_{\sigma}^{\phantom{\dagger}})$ while for the Kondo model $H_{\rm imp} + H_{\rm hyb}=J \vec{S} \cdot \vec{s}_0 + V\sum_{\sigma}f_{0,\sigma}^{\dagger}f_{0,\sigma}^{\phantom{\dagger}}$ with $\vec{s}_0=\tfrac{1}{2}\sum_{\alpha,\beta}f^{\dagger}_{0,\alpha} \vec{\sigma}_{\alpha\beta} f^{\phantom{\dagger}}_{0,\beta}$. From Eq.~\eqref{eq:Hn} we obtain the recursion relation,
\begin{equation}\label{eq:recursion}
\begin{split}
    H_{N}=&H_{N-1}+\sum_{\sigma}\Big [\epsilon_N^{\phantom{\dagger}} f_{N,\sigma}^{\dagger}f_{N,\sigma}^{\phantom{\dagger}}\\&+ t_N^{\phantom{\dagger}} \left ( f_{N-1,\sigma}^{\dagger}f_{N,\sigma}^{\phantom{\dagger}}+f_{N,\sigma}^{\dagger}f_{N-1,\sigma}^{\phantom{\dagger}}\right ) \Big ] \;,
\end{split}
\end{equation}
such that the full (discretized) model is obtained as $H^{\rm disc}=\lim_{N\to \infty} H_N$ \cite{note_nrg}. \\
(v) Starting from the impurity, we build up the chain by successively adding Wilson chain sites using the recursion, Eq.~\eqref{eq:recursion}. At each step $N$, the intermediate Hamiltonian $H_N$ is diagonalized, and only the lowest $N_s$ states are retained to construct the Hamiltonian $H_{N+1}$ at the next step (the higher energy states are discarded). In such a way, we focus on progressively lower energy scales with each iteration. Furthermore, the iterative diagonalization and truncation procedure can be viewed as an RG transformation \cite{wilson1975renormalization}, $H_{N+1}=\mathcal{R}[H_N]$.\\
(vi) The partition function $Z_N$ can be calculated from the diagonalized Hamiltonian $H_N$ at each step $N$. Wilson used RG arguments to show \cite{wilson1975renormalization} that thermodynamic properties obtained from $Z_N$ at an effective temperature $T_N \sim \Lambda^{-N/2}$ accurately approximate those of the original undiscretized model at the same temperature. The sequence of $H_N$ can therefore be viewed as coarse-grained versions of the full model, which faithfully capture the physics at progressively lower and lower temperatures.\\
(vii) The discarded states at each step form a complete set (the Anders-Schiller basis \cite{anders2006spin}), from which the NRG full density matrix can be constructed. This provides a rigorous way of calculating real-frequency dynamical quantities via the Lehmann representation \cite{weichselbaum2007sum}.

In this work, we take the DoS $\rho(\omega)$ of the TBG system for a given twist angle $\theta$ (as calculated from the BM model in Sec.~\ref{sec:BM-model}), discretize it logarithmically, and map to Wilson chains. The DoS used and the resulting Wilson chains are shown in Appendix B. NRG is then used to solve the Anderson and Kondo models describing an impurity embedded in the TBG host. Thermodynamic and dynamical quantities are calculated and discussed in Sec.~\ref{sec:Kondo-BM}. Throughout, we use an NRG discretization parameter $\Lambda=2$ and retain $N_s=4000$ states at each step of the calculation.

%##################

\subsection{Observables}
\label{sec:obs} 

 The physics of an impurity in the TBG host can be characterized by a number of observables, both thermodynamic and dynamical. 
 Here we consider the impurity contribution to the total thermal entropy $S_{\rm imp}(T)$ as a function of the temperature $T$, and the low-$T$ impurity spectral function $A(\omega)$ as a function of energy $\omega$. 
 
 The impurity entropy readily allows us to extract the Kondo scale $T_K$ from NRG data, to track accurately the RG flow, and to identify RG fixed points. It is defined as $S_{\rm{imp}}(T) = S_{\rm{tot}}(T) - S_0(T)$, where $S_{\rm{tot}}$ is the entropy of the entire system, while $S_0$ is the entropy of the free TBG host without impurities. The residual  impurity entropy $S_{\rm imp}(T=0)$, is a finite universal number of order unity which characterizes the stable RG fixed point and hence the ground state of the system; for example $S_{\rm imp}(0)=\ln 2$ for a free, unscreened impurity spin $S=\tfrac{1}{2}$ local moment.
 
 The impurity spectral function gives dynamical information and is accessible experimentally via scanning tunneling spectroscopy (STS), which probes the energy-resolved impurity density of states \cite{ternes2008spectroscopic}. For an Anderson impurity it is related to the impurity Green's function, $A(\omega)=-\tfrac{1}{\pi}{\rm Im}~G_{dd}(\omega)$ where $G_{dd}(\omega)=\langle\langle d_{\sigma}^{\phantom{\dagger}} ; d_{\sigma}^{\dagger}\rangle\rangle$. 
 
 Electronic scattering in the TBG system induced by the impurity is characterized by the t-matrix, which in turn is controlled by the impurity Green's function. In momentum space, the t-matrix equation reads,
 \begin{equation}
     G_{kk'}(\omega) = \delta_{kk'}G_{kk}^0(\omega) + G_{kk}^0(\omega)T_{kk'}(\omega) G_{k'k'}^0(\omega) \;,
 \end{equation}
where $G_{kk'}^{(0)}(\omega)=\langle\langle c_{k,\sigma}^{\phantom{\dagger}} ; c_{k',\sigma}^{\dagger}\rangle\rangle^{(0)}$ is the electron Green's function for the full (free) TBG system with (without) the impurity, and $T_{kk'}(\omega)=g^2\xi_k\xi_{k'}G_{dd}(\omega)$ is the t-matrix itself. Transforming to real space, the t-matrix equation becomes,
 \begin{equation}
     G_{\vec{r}_i\vec{r}_j}(\omega) = G_{\vec{r}_i\vec{r}_j}^0(\omega)  + G^0_{\vec{r}_i\vec{r}_0}(\omega) T(\omega) G^0_{\vec{r}_0\vec{r}_j}(\omega) \;,
 \end{equation}
where $G_{\vec{r}_i\vec{r}_j}^{(0)}(\omega)$ are the full (free) electronic propagators between real-space sites $\vec{r}_i$ and $\vec{r}_j$ of the TBG system, and the local t-matrix is $T(\omega)=g^2 G_{dd}(\omega)$. The impurity is taken to be located at site $\vec{r}_0$.

A related experimental quantity obtained by Fourier transform STS \cite{hoffman2002imaging} (FT-STS) is the quasiparticle interference (QPI) pattern, defined as,
 \begin{equation}
    \Delta \rho(\vec{q},\omega) = \sum_i e^{-i\vec{q}\cdot \vec{r}_i}\Delta \rho(\vec{r}_i,\omega) \;,
 \end{equation}
where $\Delta \rho(\vec{r}_i,\omega)=-\tfrac{1}{\pi}{\rm Im}~[G_{\vec{r}_i\vec{r}_i}(\omega)-G_{\vec{r}_i\vec{r}_i}^0(\omega)]$ is the difference in electronic density at site $\vec{r}_i$ due to the impurity. As such, the QPI pattern $\Delta \rho(\vec{q},\omega)$ can be obtained entirely from the free TBG propagators and the impurity Green's function, via the t-matrix equation.

In the following, we study the zero temperature, $T=0$, impurity spectral function $A(\omega)$ using NRG; the t-matrix and QPI can be obtained from this as described above.

%%%%%%%%%%%%%%%%%%%%%%%%%%%%%%%%%%%%%%%%%%%%%%%%%%%%%

\subsection{Limiting cases of the Kondo problem}\label{sec:KondoDOS}

The physics of the Kondo model strongly depends on the host DoS. For the problem of a magnetic impurity in a TBG host, as modelled using the BM model, there are a number of relevant limits. We assume that the impurity couples to all the orbitals in the same way meaning the local TBG DoS $\rho(\omega)$ is the only relevant quantity characterizing the host. Furthermore, we consider here the particle-hole symmetric case $V=0$ for simplicity.

The main insights of the ensuing discussion are summarized in Table~\ref{tab:ScalingTk}.
    \begin{table*}
    \centering
    \begin{tabular}{|c||c|c|c|} \hline
             & Density of States, $\rho(\omega)$   &  ~~Kondo Temperature, $T_K$~~ & ~~Impurity entropy, $S_{\rm{imp}}(T=0)$~~ \\ \hline
            Metal   & $\rho_0$ ~:~constant & $  D e^{-{\nicefrac{1}{\rho_0 J}}}$ & $0$ \\ \hline 
            Standard van-Hove  & $ \rho_0 \left[1+a\ln \left(D/|\omega| \right) \right]$ ~:~log diverging & $ D e^{\frac{1}{a} - \frac{1}{a}\sqrt{1 + \frac{2a}{\rho_0J}}}$ & $0 $ \\\hline
            Higher Order van-Hove & $ \rho_0\abs{\omega}^{-\alpha}$ ~:~power-law diverging & $D\left(1+\frac{\alpha D^\alpha}{\rho_0 J}\right)^{-\nicefrac{1}{\alpha}} $ & $-2\alpha \ln 2$\\ \hline 
            Dirac cone & $\rho_0 |\omega|$ ~:~linear pseudogap & $T_K=0$ & $\ln 2$\\ \hline
             \end{tabular}
    \caption{Summary of properties for the Kondo model with the different host DoS encountered in this work.}
    \label{tab:ScalingTk}
    \end{table*}

%##########

\subsubsection{The metallic limit}\label{subsec:SKondo}

The case of a magnetic impurity embedded in a metallic host is the most commonly encountered and well-studied situation, with extensive literature to its name (see Ref.~\cite{Hewson} for an introduction). The most important feature of the problem is that upon reducing the temperature, the system becomes increasingly strongly correlated. This is captured by the poor-man's scaling equation, Eq.~\eqref{eq:poorman}. In a metal, it is a reasonable assumption that the DoS is roughly constant within the relevant energy window, $\rho(\omega)\approx \rho_0$. Integrating  Eq.~\eqref{eq:poorman} then straightforwardly yields, 
\begin{eqnarray}
J(\Lambda)=\frac{J_0}{1+\rho_0 J_0 \ln \left( \frac{\Lambda}{D}\right)}\;,
\end{eqnarray}
where $\Lambda$ is the energy scale of interest whereas $D$ is the starting high-energy cutoff (physically $D$ is the conduction electron bandwidth). $J_0$ is the starting value of the coupling constant at the scale $D$ (with $J_0\equiv J$ in Eq.~\eqref{eq:KM}) whereas $J(\Lambda)$ is the running coupling strength at energy scale $\Lambda$. Perturbation theory breaks down, once the running coupling $J(\Lambda=T_K)\to \infty$. This happens at the root of the denominator and the corresponding running energy scale reads
\begin{eqnarray}
T_K=D e^{-1/(\rho_0 J_0)}\;.
\end{eqnarray}
This energy scale has a number of interpretations. One implication is that for temperatures $T \gg T_K$, the physics can be captured using a perturbative expansion around $J=0$, meaning the limit of a free impurity henceforth referred to as the local moment (LM) regime. Consequently, the impurity entropy, up to small correction, is that of a free spin, meaning $S_{\rm{imp}}^{\rm LM}=\log 2$. 

Conversely, for $T \sim  T_K$, perturbation theory breaks down. Nozi\`eres showed~\cite{nozieres1980kondo} that at zero temperature, the ground state is a complicated many-body spin-singlet state and the system is a local Fermi liquid. The low-temperature limit $T\ll T_K$ is referred to as the strong coupling (SC) regime.  The corresponding impurity entropy is that of a unique state with $S_{\rm{imp}}^{\rm SC}=0$. The strong coupling physics in this regime shows up in the local spectral function $A(\omega)$ as a strong quasiparticle resonance around the Fermi energy, with a spectral pinning condition satisfying the Friedel sum rule \cite{Hewson}. The RG flow between the two fixed points is relatively simple, with $J(\Lambda)$ increasing from LM to SC as the energy scale $\Lambda$ is decreased. This is illustrated in the upper panel of Fig.~\ref{fig:RGFlow}. The Kondo scale $T_K$ is the scaling invariant along this RG flow. Note that in the metallic case, the potential scattering $V$ is strictly marginal and consequently plays no role. 

%##########

\subsubsection{Close to a van Hove singularity}

This case is a twist on the usual metallic Kondo problem, with very similar physics. The poor man's scaling analysis points to an RG flow from weak to strong coupling, with the running coupling $J(\Lambda)$ again increasing from LM to SC as the energy scale $\Lambda$ is decreased (Fig.~\ref{fig:RGFlow}, upper panel). The main difference from the standard metallic case is an enhanced Kondo temperature~\cite{Gogolin1993,Zhuravlev2011,zhuravlev2018one} due to the enhanced DoS, which diverges logarithmically at low energies for standard vHs points. 
Taking the low-energy form of the DoS to be $\rho(\omega)= \rho_0 \left(1+a\ln \left(D/|\omega| \right) \right)$, we can integrate Eq.~\eqref{eq:poorman} and, as before, extract the Kondo scale from the divergence in $J(\Lambda)$. In this case we obtain,
\begin{eqnarray}
T_K=D e^{\frac{1}{a} - \frac{1}{a}\sqrt{1 + \frac{2a}{\rho_0J_0}}}\;. 
\label{eq:TKLog}
\end{eqnarray}
Note that for $a \to 0$ the DoS becomes metallic and we recover the metallic limit result for $T_K$.

This expression has two limiting cases and the relevant parameter is now $a/(\rho_0 J_0)$: (i) For $a/(\rho_0 J_0) \ll 1$, the logarithmic enhancement of the DoS plays no role and one recovers the metallic limit result, $T_K=D e^{-1/(\rho_0 J_0)}$ since the system starts out already close to the SC fixed point; (ii) In the opposite limit, $a/(\rho_0 J_0) \gg 1$, we find $T_K=D e^{- \sqrt{\frac{2}{a \rho_0J_0}}}$, which is strongly boosted relative to the metallic case.
The Kondo temperature $T_K$ is therefore enhanced in the vicinity of a vHs. As expected, the impurity entropy is quenched at low temperatures, $S_{\rm{imp}}(T\to 0)=0$, embodying Kondo singlet formation. In terms of the impurity spectral function $A(\omega)$, the quasiparticle Kondo resonance around the fermi energy is in fact suppressed logarithmically by the logarithmically diverging DoS of the free host \cite{derry2015quasiparticle}. However,  the fact that $A(\omega=0)=0$ should not in this case be interpreted as a flow towards weak coupling since the relevant quantity is rather $\rho(\omega)\times A(\omega)$, which remains finite as $|\omega|\to 0$ and satisfies a generalized Friedel sum rule \cite{logan2014common} for strong coupling physics.

%##########

\subsubsection{Close to a Higher Order van Hove singularity}

As discussed in Sec.~\ref{sec:BM-model}, a HO-vHs is characterized by a power-law divergence in the DoS. In the following discussion, we neglect the metallic background on top of which the power-law divergence sits, and take the DoS to be of the form $\rho(\omega)=\rho_0 |\omega|^{-\alpha}$ with $0<\alpha<1$. This problem falls into the class of power-law Kondo problems~\cite{Mitchell2013}. Intuitively, one expects the Kondo temperature to be further enhanced through the more strongly divergent DoS. Integrating Eq.~\eqref{eq:poorman}  with this DoS yields,
\begin{eqnarray}\label{eq:tkHO}
T_K = \frac{D}{\left(1+\frac{\alpha D^\alpha}{\rho_0 J_0}\right)^{\nicefrac{1}{\alpha}}}\;.
\end{eqnarray}
This problem also has two limits, one characterized by the local moment physics of a free impurity (LM) the other limit by a particle-hole symmetric strong coupling fixed point that we henceforth call $\rm{SSC}_{\rm{HO}}$. This RG fixed point has properties that are slightly different from the metallic strong coupling fixed point SC. It is characterized by a $T \to 0$  impurity entropy of $S_{\rm{imp}}=-2\alpha \ln 2$, which is negative. We emphasize that although the total thermodynamic entropy is of course never negative, the \emph{impurity contribution} to the total system entropy as defined in Sec.~\ref{sec:obs} can be negative, if the presence of the impurity causes dramatic changes to the host (local) electronic structure, relative to the clean host. This is precisely the case for the power-law Kondo model \cite{Mitchell2013}. 
As for the impurity spectral function, the quasiparticle Kondo resonance is suppressed by the divergent free host DoS, and we find $A(\omega) \sim |\omega|^{\alpha}$ at low energies. But again, the signature of Kondo singlet formation and strong coupling physics is that $\rho(\omega)\times A(\omega)$ remains finite as $|\omega|\to 0$, which is indeed the case for $\rm{SSC}_{\rm{HO}}$.

\begin{figure}[t]
	\includegraphics[width=0.8\linewidth]{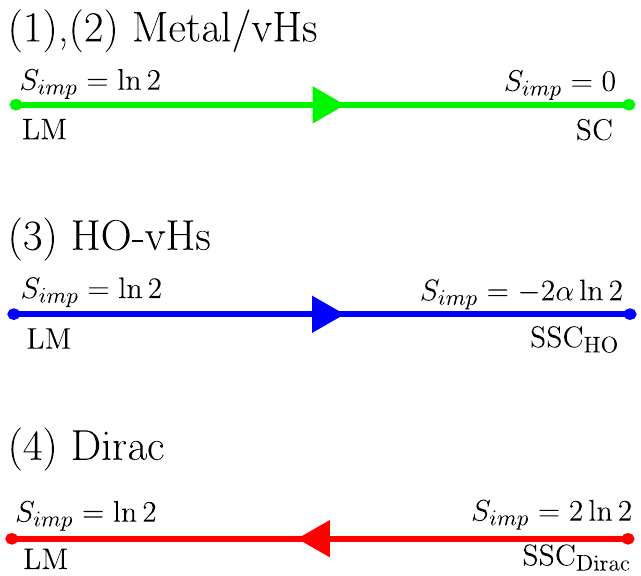}
	    \caption{
      %\justifying
      \small  One-dimensional RG diagram showing the flow into different stable fixed points as the temperature is lowered, for different host density of states profiles.}
	    \label{fig:RGFlow}
     \end{figure}

The RG flow in the particle-hole symmetric case considered here is again one-dimensional, running from LM to $\rm{SSC}_{\rm{HO}}$ as the energy scale or temperature is reduced (see middle panel of Fig.~\ref{fig:RGFlow}). We note that strong particle-hole asymmetry can play a role in the power-law Kondo model, unlike the pure metallic case, and leads to the intricate phase diagram discussed in Ref.~\cite{Mitchell2013}.

%##########

\subsubsection{The limit of a two-dimensional Dirac cone}

The linear vanishing DoS associated with a Dirac cone in two dimensional systems gives rise to a subtle impurity problem which falls into the class of so-called pseudogap Kondo models. This is by far the most complicated situation; it has been discussed at length in Refs.~\cite{Ingersent1998,Fritz2004a,Fritz2004b,Fritz2013,logan2014common}. In the following, we focus on the particle-hole symmetric case, with a linear pseudogap $\rho(\omega)=\rho_0|\omega|$ that is characteristic of 2d Dirac materials. Naively, one might expect a reduced Kondo temperature due to the reduced DoS of a Dirac cone compared with that of a metal. In fact, the Kondo effect is suppressed entirely in this case and $T_K$ vanishes, regardless of the strength of the bare coupling $J_0$. The LM fixed point is stable and the impurity local moment remains unscreened.

For large bare $J_0$ the system starts off close to the particle-hole symmetric strong coupling fixed point of the linear pseudogap Kondo model, dubbed $\rm{SSC}_{\rm{Dirac}}$. In this regime, the impurity entropy is $S_{\rm{imp}}=2 \ln 2$. However, this fixed point is unstable and RG flow on reducing the temperature tends towards the LM fixed point with entropy $S_{\rm{imp}}=\ln 2$. Unlike the other cases considered, here the renormalized running coupling $J(\Lambda)$ \emph{decreases} on reducing temperature, so that the impurity always becomes asymptotically free as $T\to 0$. This is illustrated in the lower panel of  Fig.~\ref{fig:RGFlow}. The Kondo effect can only be revived by doping so that the Dirac point is not longer at the fermi energy (in which case the low-energy DoS is finite and we recover the metallic scenario); or if very strong potential scattering is introduced.
For the linear pseudogap case $\rho(\omega)\sim |\omega|$ the impurity spectral function also goes as $A(\omega)\sim |\omega|$. Note that in this case $\rho(\omega)\times A(\omega) \to 0$ as $|\omega|\to 0$, indicating a free impurity at low energies.

%###############
%###############

\section{Results and discussion}\label{sec:Kondo-BM}
\begin{figure}[t]
	\includegraphics[width=\linewidth]{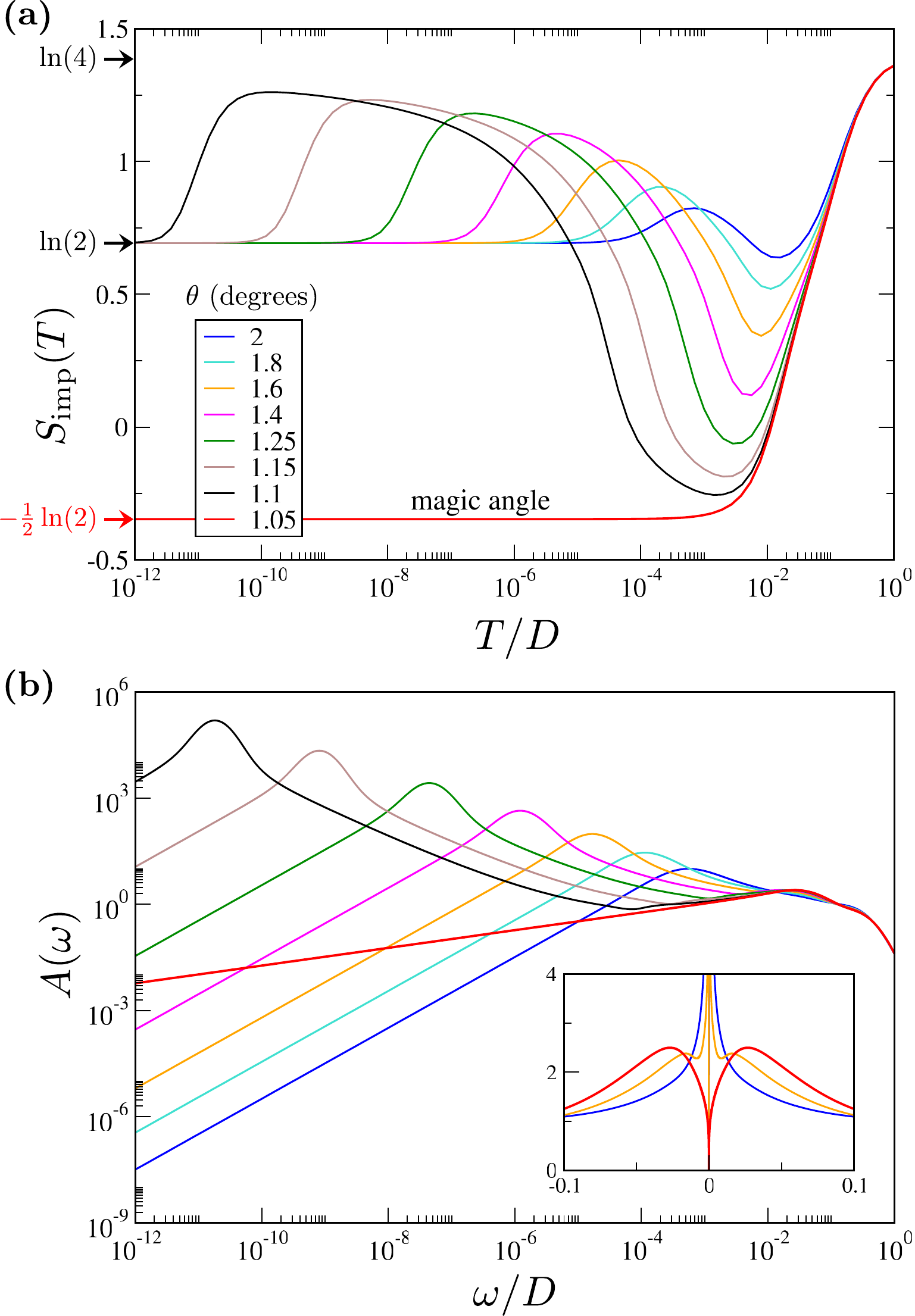}
	    \caption{
     %\justifying
     \small  NRG results for an Anderson impurity in the TBG host material described by the BM model, at various twist angles $\theta$ approaching the magic angle at $\theta=1.05^{\circ}$. 
     (a) Impurity entropy $S_{\rm imp}(T)$ vs temperature $T$; and (b) Impurity spectral function $A(\omega)$ vs energy $\omega$ at $T=0$. Inset shows low-energy spectral details on a linear scale for representative cases approaching the magic angle. All plots shown for $U_d=0.4D$, $\epsilon_d=-U_d/2$ and $g=0.2D$, with $D$ the TBG bandwidth. }
	    \label{fig:angles}
     \end{figure}

\begin{figure}[t]
	\includegraphics[width=0.8\linewidth]{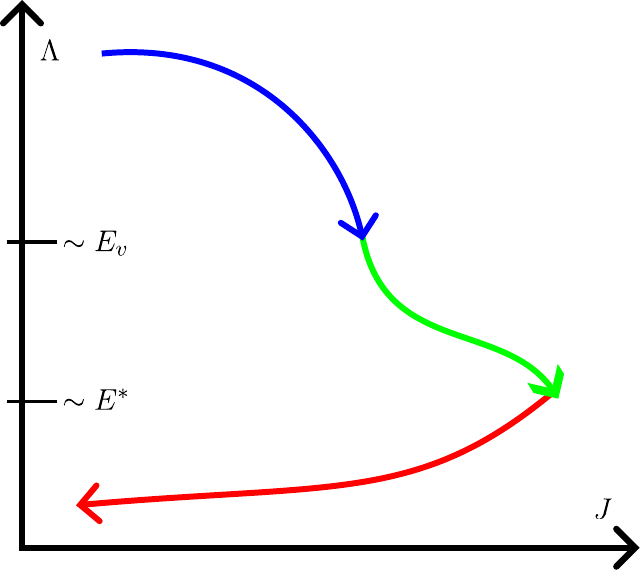}
	    \caption{
     %\justifying
     \small Near the magic angle, the TBG host DoS has a compound structure. At high energies $\Lambda$, the power-law diverging DoS associated with the HO-vHs generates a rapid RG flow towards strong coupling -- blue arrow. On the scale of $E_v$ the DoS crosses over to logarithmic vHs -- green arrow. Below $E_v$ the Dirac cone pseudogap dominates the DoS and $J(\Lambda)$ begins to decrease again as the system flows back towards the free local moment regime on an emergent scale $E^*$. The color-coding is the same as that in Fig.~\ref{fig:RGFlow}.}
	    \label{fig:flow}
     \end{figure}

\begin{figure}[t]
    \centering
    \includegraphics[width = \linewidth]{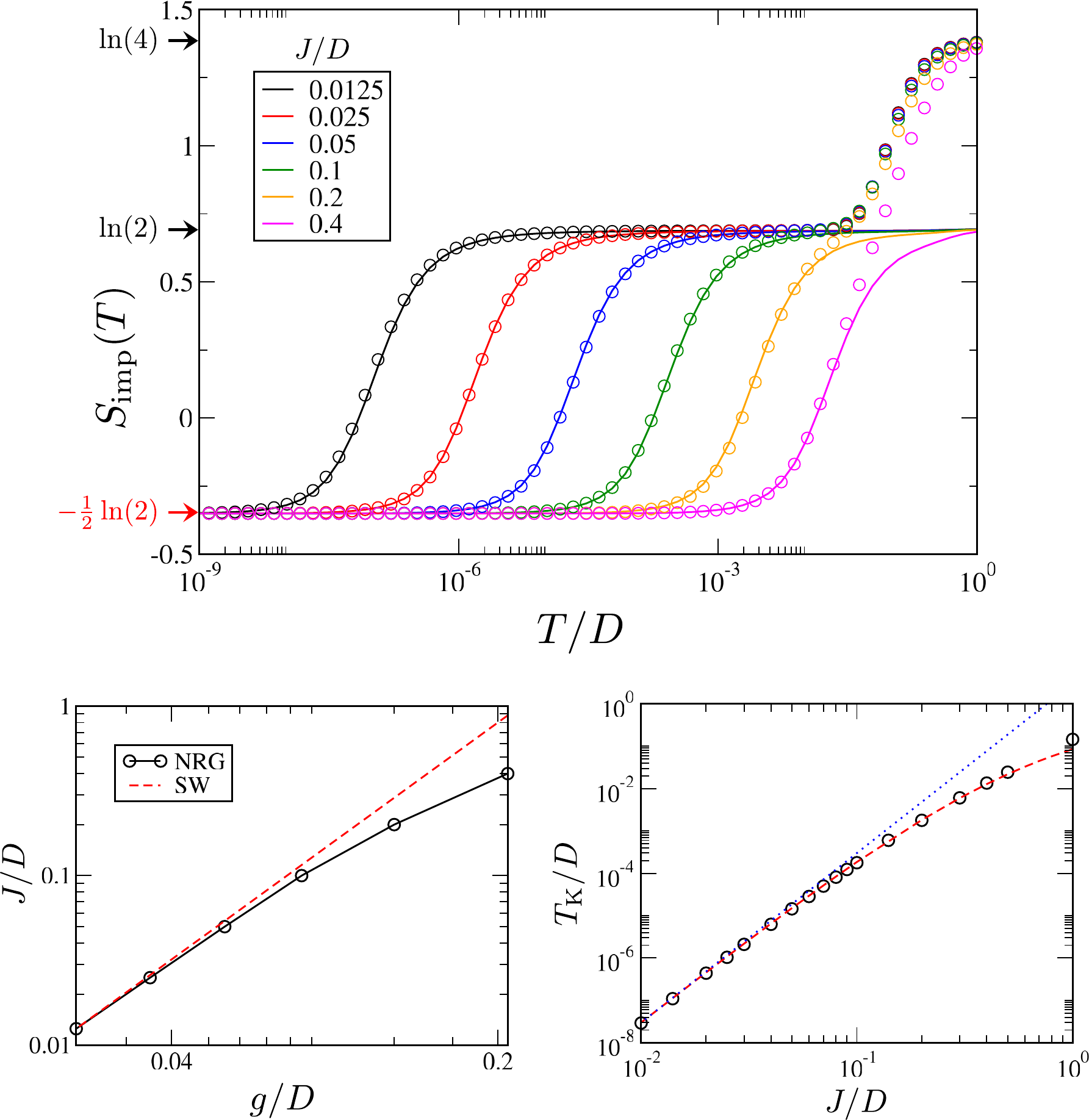}
    \caption{
    % \justifying
    \small NRG results for an impurity embedded in magic angle TBG. Main panel: Entropy $S_{\rm imp}$ vs $T$ for Anderson and Kondo impurities (points and lines respectively)  for different impurity-host couplings. For a given Kondo $J$, the low temperature physics of an Anderson model with fixed $U=0.4D$ is fit by tuning $g$. The relationship between $J$ and $g$ is shown in the lower-left inset (points), comparing with the SW result (red dashed line). The evolution of $T_K$ with $J$ is shown in the lower-right inset (points), comparing with Eq.~\ref{eq:tkHO} (red dashed line) and its small-$J$ asymptote (blue dotted line). \rev{$T_K$ is extracted numerically from NRG results for the entropy, defined in practice via $S_{\rm imp}(T=T_K)=0$.}}
   \label{fig:kondomagic}
\end{figure}

We now turn to our full NRG results for a magnetic impurity embedded in the TBG host. The impurity is taken to be of Anderson type (Eq.~\ref{eq:AM}), and the BM model is used for the host (Eq.~\ref{eq:H-TBG-symmetric}). The observables of primary interest are the temperature-dependence of the entropy $S_{\rm imp}(T)$, and the energy-resolved impurity spectral function $A(\omega)$ at $T=0$, for TBG systems with different twist angles  -- see Fig.~\ref{fig:angles}. The most dramatic changes are observed in the vicinity of the magic angle at $\theta=1.05^{\circ}$, and this is where we focus our discussion. The physical quantities we calculate reveal a complex RG flow in the system, illustrated in Fig.~\ref{fig:flow}. The physics precisely at the magic angle is also investigated in detail, and the dependence of the Kondo temperature on microscopic parameters is extracted -- see Fig.~\ref{fig:kondomagic}.

Consider first the entropy flows presented in Fig.~\ref{fig:angles}(a). At the highest temperatures $T\sim D$, the impurity has four thermally populated configurations (empty, doubly-occupied, and up/down spin states) and the entropy for all systems is therefore $\ln(4)$ in this limit. On the scale $T \sim U$ the empty and doubly-occupied impurity configurations become thermally inaccessible and only the local moment states of the impurity survive. Note that this high-$T$ charge-freezing crossover is absent in the Kondo model, which features only the two impurity spin states from the outset. Far away from the magic angle, where the low-energy TBG DoS is dominated by the linear pseudogap of the Dirac cone, the Kondo effect is inoperative and the impurity spin degrees of freedom remain unscreened down to $T=0$. The impurity entropy therefore saturates at the LM value of $\ln(2)$. This is the basic picture for the blue line in Fig.~\ref{fig:angles}(a) obtained for twist angle $\theta=2^{\circ}$. In the opposite limit, when the system is tuned to the magic angle $\theta=1.05^{\circ}$  the impurity physics is dominated by the power-law divergence in the low-energy DoS of TBG. The physics in this case is effectively that of the power-law Kondo model discussed in the previous section. Since $\rho(\omega)\sim |\omega|^{-1/4}$, the entropy saturates to $-\tfrac{1}{2}\ln(2)$ for $T\ll T_K$, where the Kondo scale $T_K$ itself is strongly enhanced. This is precisely what we observe for the red line in Fig.~\ref{fig:angles}(a).

However, the situation is much more complex for twist angles close to (but not at) the magic angle. To understand the full RG flow in this intermediate regime, consider Fig.~\ref{fig:flow}, together with the fixed point discussion in Sec.~\ref{sec:KondoDOS} and the RG flow diagrams in Fig.~\ref{fig:RGFlow}. Note that the same color-coding is used in Figs.~\ref{fig:RGFlow} and \ref{fig:flow}.

Near the magic angle, the TBG host DoS has a compound structure featuring multiple elements, each of which corresponding to a different limiting Kondo problem. At high energies, the system shows behavior that is characteristic of the HO-vHs, denoted in blue in Fig.~\ref{fig:flow}. This behavior crosses over to that of a standard logarithmic vHs on the scale of $E_v$, denoted in green. Far below this scale, the linear vanishing pseudogap DoS of the Dirac cone emerges, denoted in red. Depending on the energy window, the RG flow will therefore be controlled by the different regimes depicted in  Fig.~\ref{fig:RGFlow}.

This RG flow is reflected in the temperature dependence of the entropy. After the charge degrees of freedom are frozen out on the high-temperature scale of $T\sim U$ and the impurity entropy reaches $\sim \ln(2)$ characteristic of the LM regime, the system then rapidly flows towards $\rm{SSC}_{\rm{HO}}$ on further reducing the temperature. On this trajectory, the effective coupling strength $J(\Lambda)$ grows as the energy scale $\Lambda$ decreases, and the entropy approaches $S_{\rm{imp}}=-\tfrac{1}{2} \ln 2$. 
However, at the scale $E_v$ the physics of the logarithmic vHs takes over and the system starts  to flow towards the regular SC fixed point with $S_{\rm{imp}}=0$. The running coupling $J(\Lambda)$ continues to increases. 
But on the `other side' of the vHs in energy space, as the temperature is further decreased, the effect of the low-energy pseudogap DoS begins to dominate. Interestingly though, at this point in the RG flow the system already has a very strong coupling strength $J(\Lambda)$, which puts the system close to the unstable $\rm{SSC}_{\rm{Dirac}}$ fixed point. The entropy therefore `overshoots' up to $\ln(4)$ characteristic of this fixed point. The ultimate RG flow on the lowest energy scales is therefore between $\rm{SSC}_{\rm{Dirac}}$ and the stable LM fixed point of the pseudogap Kondo problem, with a residual $T=0$ entropy of $\ln(2)$. The ground state is an unscreened local moment with $\ln(2)$ entropy in all cases except when precisely at the magic angle (where $E_v = 0$ such that this final part of the flow towards LM is omitted). Our NRG results for the entropy show that the final low-temperature flow between $\rm{SSC}_{\rm{Dirac}}$ and LM is controlled by an emergent energy scale $E^* \sim |E_v|^3$.
As we get closer to the magic angle, the $E_v$ scale reduces and the lines fold progressively onto that of the red line for the magic angle itself. The $E^*$ scale rapidly becomes very small. This gives a finite window in twist angle over which magic angle physics can be observed at intermediate temperatures.

The same RG flow is demonstrated by the $T=0$ spectral function for the impurity $A(\omega)$, which we plot in Fig.~\ref{fig:angles}(b) for the same systems. On the lowest energy scales $|\omega|\ll E^*$, we find $A(\omega)\sim |\omega|$ characteristic of the linear pseudogap Kondo model, for all cases except when precisely at the magic angle. This is because the physics here is controlled by the Dirac cone and the resulting RG flow toward the LM fixed point. By contrast, at the magic angle, the enhanced DoS leads to strong coupling physics and a flow towards the $\rm{SSC}_{\rm{HO}}$ fixed point for all $|\omega| \ll T_K$, yielding $A(\omega)\sim |\omega|^{1/4}$ (red line). As the magic angle is approached, the $E_v$ scale diminishes and so the spectrum progressively folds onto the magic angle result, see e.g.~black line for $\theta=1.1^{\circ}$ in Fig.~\ref{fig:angles}(b). The most prominent feature of the impurity spectral function is however the dramatic peak on the scale of $E^*$ (note the log scale), which characterizes the flow between $\rm{SSC}_{\rm{Dirac}}$ and LM fixed points. This is highlighted in the inset to Fig.~\ref{fig:angles}(b) which compares on a linear scale the magic angle result (red line) to systems at $\theta=1.6^{\circ}$ (orange) and $2^{\circ}$ (blue). The rapid change in position and intensity of this spectral peak on nearing the magic angle demonstrates that quantum impurities are highly sensitive probes of magic angle physics in TBG systems.

In Fig.~\ref{fig:kondomagic} we turn to an analysis of NRG results for systems at the magic angle itself. In the main panel (top) we plot the impurity entropy $S_{\rm imp}(T)$ for impurities of either Anderson type (Eq.~\ref{eq:AM}, points) or Kondo type (Eq.~\ref{eq:KM}, lines), for different impurity-host couplings. 

At the highest temperatures $T\sim D$, the Anderson impurity again shows $\ln(4)$ entropy for the four quasi-degenerate impurity states. At lower temperatures $T\sim U$, the charge configurations on the impurity in the Anderson model are frozen out and only the local moment spin states survive, giving $\ln(2)$ entropy. In this regime the system is close to the LM fixed point. RG flow towards the ${\rm SSC}_{\rm HO}$ strong coupling fixed point results in a crossover in the entropy on the scale of the Kondo temperature $T\sim T_K$ to $S_{\rm imp}=-\tfrac{1}{2}\ln(2)$. This remains the $T=0$ residual impurity entropy for TBG systems at the magic angle. However, the Kondo scale itself varies with the impurity-host coupling, as seen in the main panel of Fig.~\ref{fig:kondomagic} by the evolution of the different lines. For good scale separation $T_K\ll U$, we see clear two-stage behavior, with distinct crossovers to and from the LM fixed point in the Anderson model. However, given the strongly enhanced $T_K$ at the magic angle, such a scale separation may not be in evidence in practice (see e.g.~pink and orange lines in Fig.~\ref{fig:kondomagic} which show a more or less direct crossover in the entropy from $\ln(4)$ to $-\tfrac{1}{2}\ln(2)$; or indeed the cases close to the magic angle in Fig.~\ref{fig:angles}).

By contrast, the Kondo impurity features only the local moment spin configurations and hence has a $\ln(2)$ entropy at high temperatures $T\sim D$. The Kondo scale generated by finite antiferromagnetic exchange coupling $J$ results in the same crossover to the ${\rm SSC}_{\rm HO}$ fixed point, with the same $T=0$ residual entropy of $-\tfrac{1}{2}\ln(2)$. Indeed, RG arguments imply~\cite{Hewson} that the physics of the Anderson and Kondo models for $T\ll U$ should be identical, providing the effective Kondo coupling $J$ is chosen appropriately for a given $U$ and $g$ of the Anderson model. To verify this Anderson-Kondo mapping in the magic-angle TBG setting, in Fig.~\ref{fig:kondomagic} we considered Kondo models with different $J$ and then fit Anderson models to match the low-temperature physics by tuning $g$ at fixed $U$. In such a way, the Kondo and Anderson models have the same Kondo temperature $T_K$.  The precise agreement in the universal regime confirms that at particle-hole symmetry the effective Kondo model is a faithful description of the more microscopic Anderson model. 

The Anderson-Kondo mapping can be performed perturbatively via the approximate SW transformation \cite{Hewson,schrieffer1966relation} as described in Sec.~\ref{sec:Kondo}. 
The exact relationship between $J$ and $g$ as extracted from our NRG results is shown in the lower left panel of Fig.~\ref{fig:kondomagic} as the circle points. The SW result (red dashed line) is seen to work well when the bare coupling of the underlying Anderson model is small, $g\ll U$  (small $T_K$ regime). Away from this limit, NRG results show that the Kondo model is still the correct low-energy effective model, but that non-perturbative techniques must be used to obtain the correct effective model parameters~\cite{rigo2020machine}. The evolution of the numerically-extracted Kondo temperature as a function of the effective $J$ is shown in the lower right panel of Fig.~\ref{fig:kondomagic} (points), and is compared with the analytic result for the Kondo model Eq.~\ref{eq:tkHO}
(red dashed line). The blue dotted line is the asymptotic small-$J$ limit of this expression, $T_K \sim (4\rho_0 J)^{1/\alpha}$.

Our full NRG results for an impurity in magic angle TBG therefore confirm the analytic predictions of the previous sections. For a comparison with results for an impurity coupled to a standard logarithmic vHs, see Appendix~C.

\rev{Finally, we comment on the role of potential scattering and particle-hole symmetry breaking. In the above analysis we have for simplicity neglected particle-hole asymmetry in the TBG host DoS by employing the symmetric BM model. However, we believe this approximation is well-justified and does not affect the presented results. Although in principle particle-hole asymmetry can lead to Kondo screening in the linear pseudogap case \cite{Fritz2013} relevant to the low-energy Dirac cone in TBG away from the magic angle, the singlet-doublet quantum phase transition arises only at very strong asymmetry. In practice, the relatively small particle-hole symmetry breaking in TBG means that the impurity problem is far away from the asymmetric strong coupling Kondo phase. Within the doublet local moment phase, particle-hole asymmetry is RG irrelevant and can be safely ignored. We have also assumed that the impurity itself is particle-hole symmetric ($\epsilon_d=-U_d/2$ in the Anderson model, or $V=0$ in the Kondo model). Relaxing this condition induces potential scattering in the TBG host. Very strong deviations away from the half-filled Anderson impurity are required to destroy the local moment ground state (the resulting asymmetric Kondo strong coupling state is continuously connected to the trivial empty orbital state of the impurity). In this regime, the mapping to the Kondo model breaks down (the large value of $V$ in the Kondo model required for Kondo screening is unphysical). Therefore, we argue that the results presented above are generic for a local moment impurity embedded in a TBG host material.}

%############

\section{Conclusions}
\label{sec:conclusions}
In this paper, we have studied the physics of a single magnetic impurity in TBG at, and close to, the magic angle. We find a surprisingly rich range of behavior, rooted in the unique evolution of the TBG density of states. It is interesting to note that there is no Kondo screened ground state in general, only at the magic angle. However, the signatures at finite temperature relevant to experiment show highly nontrivial structure due to the interplay between van Hove and Dirac physics on the level of a strongly correlated quantum impurity problem. Close to the magic angle, the TBG host density of states at different energy scales yields different limits of paradigmatic Kondo models -- from logarithmic and power-law diverging Kondo to pseudogap vanishing Kondo. The subtle renormalization group flow between these limits shows up in the temperature and energy dependence of physical observables.

The behavior we uncover should be detectable in STM experiments. Indeed we argue that the impurity response in TBG gives a very clear signature of magic angle physics. Magnetic impurities may therefore prove useful as highly sensitive \textit{in-situ} probes for moir\'{e} materials.

An interesting direction of future research is the role of the RKKY interaction between multiple magnetic impurities in TBG, and how it competes with the Kondo effect of individual impurities near the magic angle.

\rev{Although van Hove-boosted Kondo physics may be observable in other systems (including 3d bulk metals \cite{igoshev2019giant,igoshev2022giant} with magnetic impurities), we note that TBG stands out as a uniquely tunable platform. Furthermore, TBG also allows one to study the complex interplay of these effects with Dirac physics.}

%%%%%%%%%%%%%%%

\begin{acknowledgments}
We thank S.~Polla, V.~Cheianov, L.~Classen, A.~Chubukov, and L.~Fu for helpful discussions. this work is part of the D-ITP consortium,
a program of the Netherlands Organisation for Scientific Research (NWO) that is funded by the Dutch Ministry of Education, Culture and Science (OCW). DOO acknowledges the support from the Netherlands Organization for Scientific Research (NWO/OCW) and from the European Research Council (ERC) under the European Union's Horizon 2020 research and innovation programme. AKM acknowledges funding from the Irish Research Council through the Laureate Award 2017/2018 grant IRCLA/2017/169.
\end{acknowledgments}

%##################

\appendix

\section{Locating saddle point positions in the particle-hole symmetric BM model}

To locate the saddle points in the Brillouin  zone, we require gradients of the energy. The flat bands make it imperative that the first derivative is computed very accurately, and this means that we must avoid crude finite difference methods for numerical derivatives of the energy. This problem can be addressed by utilizing our analytical access to the Hamiltonian itself, and implementing the Hellman-Feynman theorem,
\begin{equation}
    \pdv{E}{k^i} = \expval{\pdv{H}{k^i}}{\psi_k}, 
\end{equation}
where $\psi_k$ is the wavefunction of the lowest positive energy band at momentum $\vec{k}$. A more extended analysis of this method has been introduced recently in \cite{chandrasekaran2022detect}.

The saddle point is found by minimizing the value of $\left(\partial_{k_x} \epsilon_\vec{k}\right)^2 + \left(\partial_{k_y} \epsilon_\vec{k}\right)^2$ over the Brillouin  zone vectors, by using the conjugate gradient method.
The coefficients for the local energy dispersion are then calculated along the principal directions of the saddle point. These directions are given by the eigenvectors of the Hessian matrix, and are in general different from the $k_x$ and $k_y$ of the Brillouin  zone. Derivatives are convenient to compute along the natural Brillouin  zone directions, however. The trick to overcome this is to note that a matrix formed from $n$ derivatives of a scalar transforms like a rank-$n$ tensor. Then the tensor transformation rule can be used with a covariant Jacobian to obtain the coefficients along the rotated axes. 

One subtlety is that for angles very close to the magic angle, there are secondary vHs points at other locations in the Brillouin zone than those indicated in Fig.~\ref{fig:contourslog}. For the purposes of the Kondo effect, we have only considered the one with the largest spectral weight, since this is found to dominate the results of our NRG calculations. This is done by computing the local dispersion coefficients $\alpha$ and $\beta$ at each of the saddle points, and picking the one with the largest value of $\frac{1}{\sqrt{\alpha\beta}}$. 

%!!!!!!!!!!!!

\begin{figure}[t!]
    \centering
    \includegraphics[width = \linewidth]{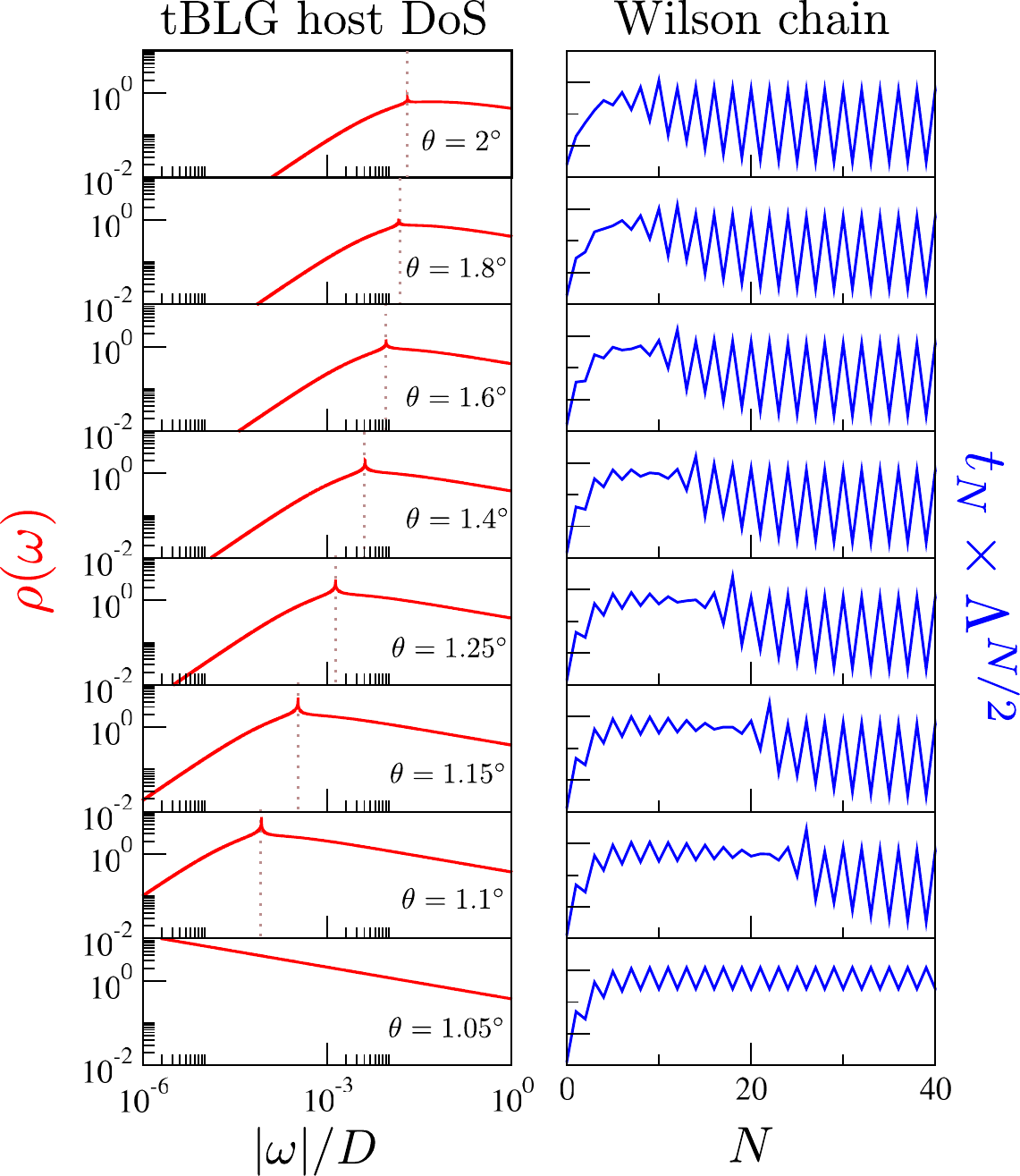}
    \caption{
    %\justifying
    \small Left panels: DoS used in our NRG calculations, obtained from analysis of the effective BM model, for different twist angles $\theta$ as used in the main text. Vertical dotted lines show the $E_v$ scale at which the vHs divergence occurs. This scale moves to lower energies as the magic angle is approached. Right panels: corresponding Wilson chain coefficients.}
    \label{fig:wc}
\end{figure}

\begin{figure}[h!]
    \centering
    \includegraphics[width = \linewidth]{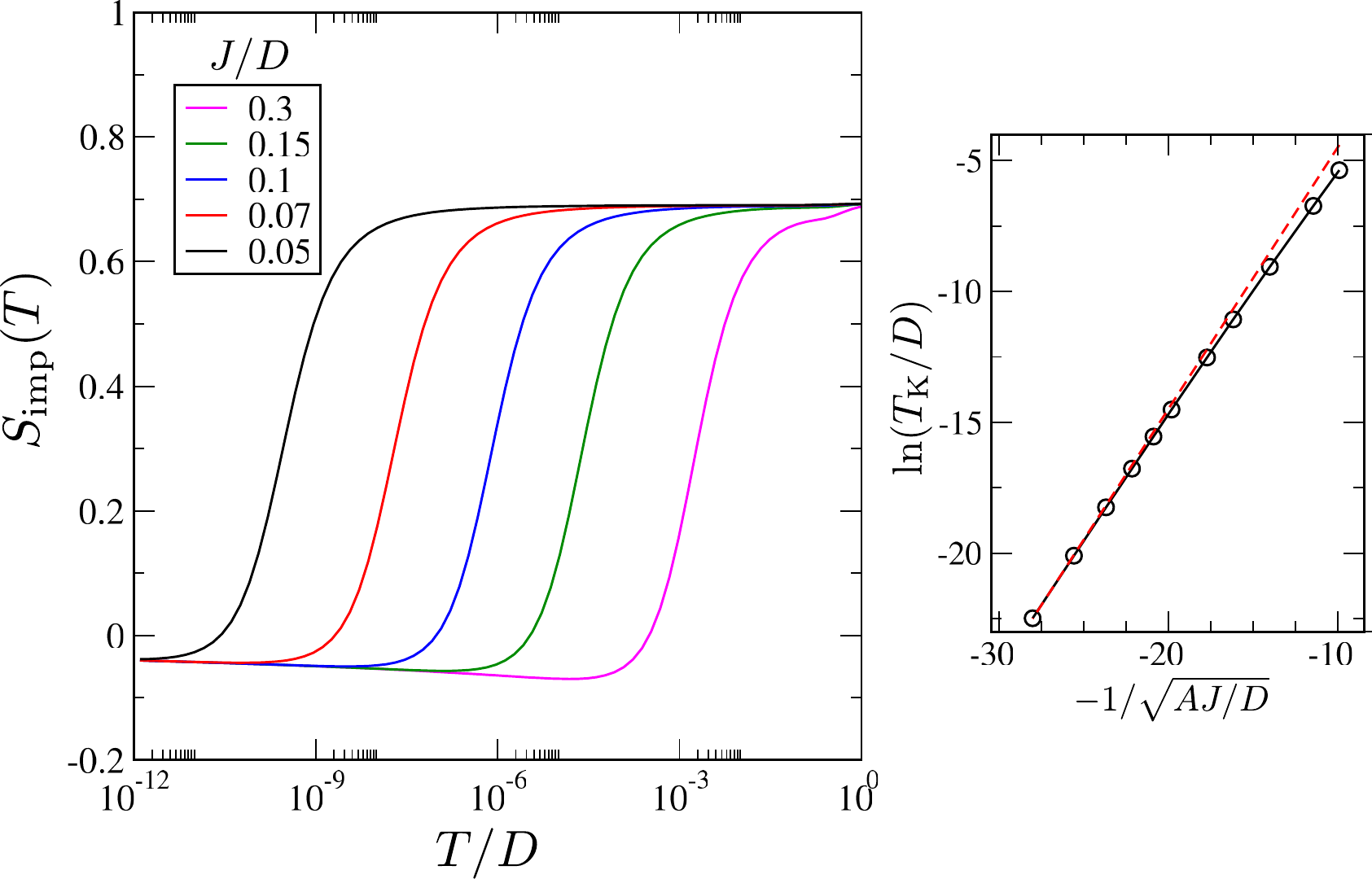}
    \caption{
   % \justifying
   \small Main panel: NRG results for the impurity entropy $S_{\rm imp}$ vs temperature $T$ for different bare coupling strengths $J$, in a system with a pure log-diverging DoS (standard vHs). Right inset shows the extracted $T_K$ scale (points), compared with Eq.~\ref{eq:TKLog} (black line) and the asymptotic result $T_K \sim D e^{-1/\sqrt{A J/D}}$ (red dashed line), with $A=a\rho_0 D/2$.}
    \label{fig:Slog}
\end{figure}

%###############

\section{Mapping from TBG density of states to the NRG Wilson chain}

The TBG DoS away from the magic angle has two qualitative features: the linear-pseudogap Dirac cone at low energies and the divergence due to the vHs on the scale of $E_v$. We extract an effective model DoS from analysis of the BM model for different twist angles -- see left panels of Fig.~\ref{fig:wc} for the cases explicitly considered in the main text (we have rescaled the energy range in terms of the bandwidth cutoff and normalized the spectrum to unity). This DoS is then discretized logarithmically and mapped to a Wilson chain \cite{bulla2008numerical} as described in Sec.~\ref{subsec:nrg}. The corresponding Wilson chain hopping parameters are plotted in the right panels of Fig.~\ref{fig:wc}. The results show that the different DoS elements can be captured in NRG through the crossover behavior in the functional form of the Wilson chain.

%##################

\section{NRG calculations for an impurity coupled to a conventional log-vHs host}

In Fig.~\ref{fig:Slog} we provide reference NRG calculations for a Kondo impurity coupled to a pure log-diverging DoS. This gives a useful comparison to our result for an impurity embedded in the magic-angle TBG system, which has a HO-vHs point and hence a stronger power-law diverging DoS. As predicted from our perturbative scaling (poor man's scaling) results, the system flows towards strong coupling in all cases, in which the impurity is Kondo-screened. The residual entropy at $T=0$ is seen to be $S_{\rm imp}=0$, although this limit is approached logarithmically slowly from below. This is characteristic of the logarithmic DoS. From the scaling with bare coupling strength $J$ in the left panel, the Kondo temperature $T_K$ is seen to be enhanced relative to the metallic case, but substantially suppressed relative to the power-law diverging DoS case. In the right panel, we analyze the behavior of $T_K$ in more detail, comparing NRG results at different $J$ (circle points) with our analytic formula Eq.~\ref{eq:TKLog} (black line). The results agree almost perfectly. The red dashed line is the asymptotic result at small $J$, which also does remarkably well compared with exact NRG results.

%####################
	
\bibliography{biblio}
	
\end{document}